\documentclass[pre,preprintnumbers,amsmath,amssymb,superscriptaddress]{revtex4}
\usepackage{amsmath}
\usepackage{mathtools}
\usepackage{graphicx}
\usepackage{dcolumn}
\usepackage{booktabs}
\usepackage{bm}
\usepackage{longtable}

\DeclareMathOperator{\less}{<}
\DeclareMathOperator{\great}{>}
\usepackage{hyperref}
\usepackage{multirow}
\hypersetup{
	colorlinks=true,
	linkcolor=blue,
	citecolor=blue,
	urlcolor=blue
}
\usepackage{epstopdf}
\allowdisplaybreaks

\begin{document}

\preprint{APS/123-QED}

\title{Rare Events and Single Big Jump Effects in Ornstein-Uhlenbeck Processes}

\author{Alberto Bassanoni}
\affiliation{Dipartimento di Scienze Matematiche, Fisiche ed Informatiche,
Università degli Studi di Parma, Parco Area delle Scienze 7/A, 43124, 
Parma, Italy}
\affiliation{INFN, Gruppo Collegato di Parma, Università degli Studi di Parma, Parco Area delle Scienze 7/A, 43124, Parma, Italy}

\author{Alessandro Vezzani}
\affiliation{Dipartimento di Scienze Matematiche, Fisiche ed Informatiche, 
Università degli Studi di Parma, Parco Area delle Scienze 7/A, 43124, 
Parma, Italy}
\affiliation{IMEM, CNR Parco Area delle Scienze 37/A 43124 Parma}

\author{Eli Barkai}
\affiliation{Department of Physics, Institute of Nanotechnology and Advanced Materials, Bar-Ilan University, Ramat-Gan, 52900, Israel}

\author{Raffaella Burioni}
\affiliation{Dipartimento di Scienze Matematiche, Fisiche ed Informatiche,
Università degli Studi di Parma, Parco Area delle Scienze 7/A, 43124, 
Parma, Italy}
\affiliation{INFN, Gruppo Collegato di Parma, Università degli Studi di Parma, Parco Area delle Scienze 7/A, 43124, Parma, Italy}

\date{\today}

\begin{abstract}
\noindent 
Even in a simple stochastic process, the study of the full distribution of time integrated observables can be a difficult task. This is the case of a much-studied process such as the Ornstein-Uhlenbeck process where, recently, anomalous dynamical scaling of large deviations of time integrated functionals has been highlighted. Using the mapping of a continuous stochastic process to a continuous time random walk via the ``excursions technique'', we introduce a comprehensive formalism that enables the calculation of the complete distribution of the time-integrated observable $A = \int_0^T v^n(t) dt$, where $n$ is a positive integer and $v(t)$ is the random velocity of a particle following Ornstein-Uhlenbeck dynamics.  We reveal an interesting connection between the anomalous rate function associated with the observable $A$ and the statistics of the area under the first-passage functional during an excursion. The rate function of the latter, analyzed here for the first time, exhibits anomalous scaling behavior and a critical point in its dynamics, both of which are explored in detail. The case of the anomalous scaling of large deviations, originally associated to the presence of an instantonic solution in the weak noise regime of a path integral approach, is here produced by a so called ``big jump effect'', in which the contribution to rare events is dominated by the largest excursion. Our approach, which is quite general for continuous stochastic processes, allows to associate a physical meaning to the anomalous scaling of large deviations, through the big jump principle. 
\end{abstract}

\maketitle

\section{Introduction}

\noindent The study of rare events in stochastic processes is an important cross-cutting topic covering areas of mathematics \cite{GumbelExtremeIntro, FrechetExtremeIntro}, physics \cite{VulpianiIntro, HollanderIntro}, geophysics \cite{DeArcangelis_Intro_Geo}, economics \cite{BlackScholesIntro, Filiasi_Intro_Fin, EmbrechtsIntro} and social sciences \cite{AlbeverioIntro}.  In all these areas, it is of fundamental importance to be able to correctly assess the probability that an event belongs to the tail of a distribution. Very powerful techniques have been developed for this purpose, the best known and most applied of which is the theory of large deviations \cite{TouchetteLDT, TouchetteLDT2, Baek_Sing_LDT, VivoLDT,DemboLDT, MajumdarLDT}. This is based on the construction of the so-called rate function and basically captures rare events which are exponentially suppressed as a function  of some parameter, such as the number of random components of a system, the time of observation of a stochastic process, or the amplitude of the noise acting on a dynamical system. It is quite natural to think that these rare, exponentially suppressed events are produced by many small fluctuations, all coherent, all contributing in the same direction and thus adding up to a large fluctuation. 

However, this is not the sole mechanism by which rare events can be generated. An alternative process for triggering rare events that cannot simply be traced back to the construction of the rate function, the so-called big jump principle \cite{Chistyakov_BJ, BJFoss}, has attracted much interest. In this alternative regime, rare events are generated by the essential contribution of a single event that overpowers all other fluctuations and dominates the feature of the tail distribution, and is typical of processes with subexponential tails. The big jump principle has been used in recent studies to estimate the probabilities of rare events in various physical systems where traditional large deviation theory proves inadequate. This principle finds applications in anomalous transport contexts, such as condensation phenomena in various microscopic models \cite{Condensation_Mori, Condensation_Corberi, Condensation_Gradenigo, Condensation_Maj_Zia, Condensation_MezardEconomy}, active matter \cite{Active_matter_1, Active_matter_2, Active_matter_3, Active_matter_4, Active_matter_5}, random matrix theory \cite{Random_matrix_1, Random_matrix_2, Random_matrix_3} or stochastic resetting processes \cite{Stochastic_resetting_1, Stochastic_resetting_2, Stochastic_resetting_3}, typically in the presence of L\'evy statistics and power law distributions \cite{BJ1, BJ2, BJ3, BJ4, BJ5, BJ_sub_exp, BJexit, paperEVS, LevyLorentz1, LevyLorentz2, Artuso_LevyLorentz, LevyLorentzBarkai, BJ_Holl, LevyLorentzBeenakker}. Although it is not yet very clear to which classes of systems the principle is applicable, its validity appears to be very broad and covers all processes in which it is possible to identify a class of independent ‘jumps’ (renewals) that follow a subexponential distribution and can cause the rare event \cite{BarkaiLaplaceErrors}. 

A notable challenge resides in establishing the presence and function of the single big jump mechanism for continuous time correlated stochastic processes. An optimal experimental framework to investigate this phenomenon is provided by the Ornstein-Uhlenbeck process, one of the fundamental models in physics which characterizes the velocity dynamics of a Brownian particle or an overdamped Langevin process within a harmonic potential. This is one of the stochastic processes most studied in probability theory \cite{Karatzas}, with wide applications in physics \cite{Barkai_OU_Phys}, in finance \cite{Barndorff_Ou_Fin, ChicheBouch14_Fin} and also in climate science \cite{OU_Intro_Climate1, OU_Intro_Climate2}. It is a Markovian, ergodic, homogeneous and non-critical process with a Gaussian stationary probability distribution for its position at any given time. As a result, its rare events statistics fall within the standard theory of large deviation. However, this is no longer true if one considers the probability distribution of more complex dynamical observables. Specifically, recent studies have focused on the rare events of a time-integrated observable dependent on the stochastic velocity of a particle subject to overdamped Langevin dynamics, described by an Ornstein-Uhlenbeck process. These studies have revealed that the rare events of the probability density function (PDF) $P(A,T)$ of the observable $A = \int_0^T v^n(t) dt$ at a specified time $T$ exhibit an anomalous sub-exponential stretched behavior, when $n>2$.
The result was obtained by extending the standard theory of large deviations and the calculation of the rate function to the case where the decay of the PDF is not exponential in the parameter considered (here time) but follows a sub-exponential decay, resulting in an anomalous scaling of dynamical large deviations. So in this case the standard large deviations theory works, but in an anomalous scaling limit. This was found by a path integral approach \cite{TouchetteInst, MeersonInst, NaftaliInst, HarrisInst, AlqahtaniInst, ChenInst, Meerson2Inst} where, in the weak noise regime, the emergence of a stable instantonic solution is the source of the anomalous scaling. The same technique has been used to study rare events of the observable $A$ in the $n=1$ case of one-dimensional Brownian motion $x(t)$ in a trapping potential $\sim |x|$, and similarly anomalous large deviations are observed \cite{SmithSwept_Pot}.

At this point, an interesting question is whether the anomalous scaling of large deviations coincides with big jump regimes. To answer this question, we must be able to identify the event corresponding to the big jump in continuous processes, and to do this, our aim is to find a well-defined mapping to describe the continuous-time process in terms of a continuous time random walk (CTRW) \cite{Barkai1, Barkai2, Barkai3}. We implement this mapping using the "excursion technique", which allows us to formulate a continuous process in terms of a CTRW. Using the mathematical tools of first passage problems \cite{Redner, Gardiner, VanKampen, Martingales_FPA}, we reinterpret the continuous process as a discrete renewal process around the zero line by counting the number of times its trajectory hits the zero line in a finite time frame. This renewal process is made of a finite and numerable independent and identically distributed (IID) random variables and it can be defined for any finite observation time. If in this new CTRW the displacements are drawn from a probability distribution with sub-exponential tails, then it is possible to apply the single big jump principle to characterize its rare event statistics.
 
We study in detail our original process with the CTRW obtained with the "excursion technique" and
we investigate the probability distribution of the area under a single excursion $\cal A$ as a function of the initial velocity $v_0$. As a first result, we use the CTRW approach to determine the typical fluctuations of the observable $A$, by means of an Einstein relation.
Then we show that the PDF of the excursion areas is also described by an anomalous rate function, which obeys a differential equation that displays a critical point, which we analyze in the tails. Interestingly, at large $\cal A$ the PDF is sub-exponential, and this allows us to apply the big jump principle to study rare events  and determine the tails of the PDF $P(A,T)$. This allow us to gain interesting physical insights into the behaviour of the process and link our results for the first-passage PDF with the dynamical phase transition of the global $P(A, T)$ PDF, first observed in \cite{NaftaliInst}. In particular, we connect the presence of the instantonic solution with the big jump effects in the rare events of the PDF of the time integrated observable, highlighting the formal analogy between the two approaches and giving a physical explanation to the emergence of this anomalous scaling. Our analytical predictions are tested and confirmed by extensive numerical simulations.

The paper is organized as follows: We start in Section \ref{Section2} with a description of the model and the first-passage methods we use, highlighting the main results we obtain in subsection \ref{sub_mainresults}. In Section \ref{Section3} we explore the bulk of our PDF, calculating typical fluctuations of $P(A, T)$ through the CTRW mapping. In Section \ref{Section4} we focus on the study of the distribution of the area under the first passage excursion, while in Section \ref{Section5} we make use of all the previous collected results to characterize the rare events of $P(A, T)$ via the single big jump principle. Finally, in Section \ref{Section6} we discuss about our methods and results, their possible future perspectives and open questions.

\section{Model, Methods and Main Results}
\label{Section2}

\begin{figure}
\includegraphics[scale = 0.6]{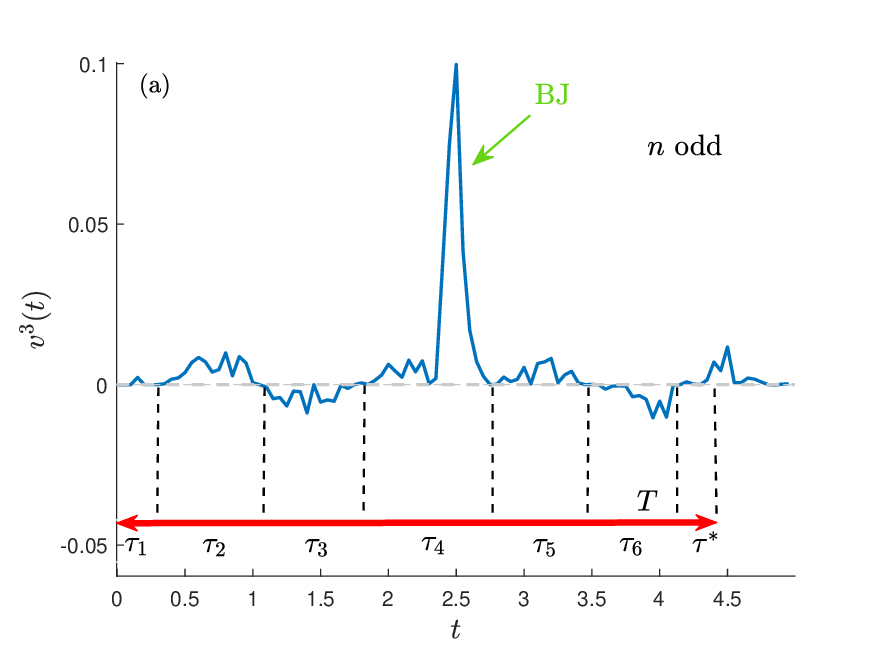}
\includegraphics[scale = 0.6]{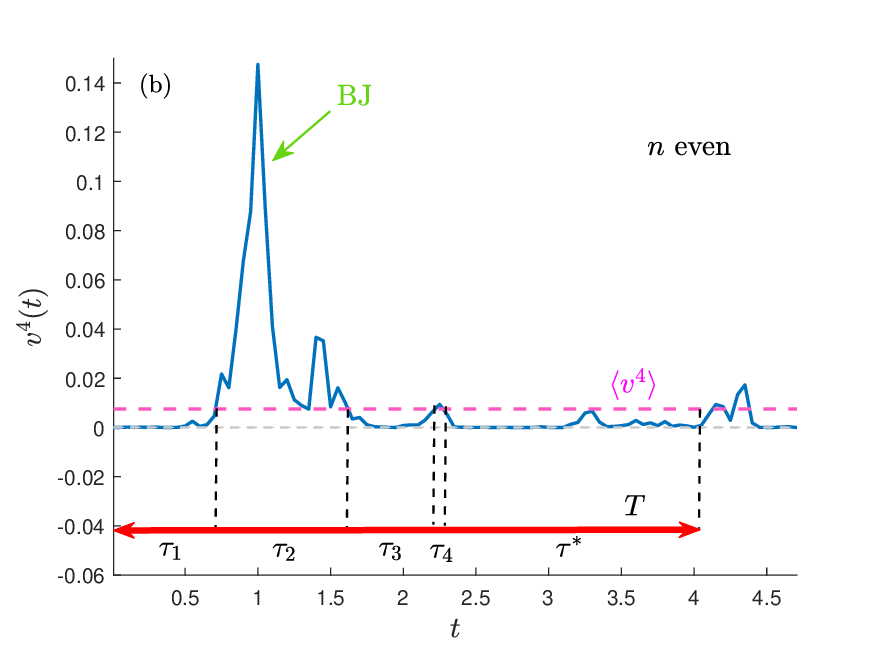}
\caption{A realization of a Langevin path for the process $v^n(t)$ governed by equation \eqref{OUequation}. In Figure (a) is shown an example of $n$ odd with $n=3$, in which the steps of the CTRW are computed as a renewal process on the zero line, represented by the dotted gray line. In Figure (b), an example of even $n$ process with $n=4$, in which the steps of the CTRW are computed as a renewal process on the equilibrium mean value $\langle v^n \rangle$, represented by the dotted pink line. For both paths the red double arrow represents the total measurement time $T$ and below  the relative renewal times $\{ \tau_1, \tau_2, ..., \tau^* \}$ are shown. The big jump (BJ), i.e., the largest area of the path that controls the rare events dynamics, is indicated with a green arrow;}
\label{fig:CTRW}
\end{figure}

\noindent 
Consider a particle of random velocity in one dimension $v(t)$ governed by an Ornstein-Uhlenbeck process, described by the following Langevin equation:

\begin{equation}
\label{OUequation}
\dot{v}(t) = -\gamma v(t) + \sigma \eta(t)
\end{equation} 
where $\gamma$ is the damping coefficient of the friction force, $\sigma$ is the noise amplitude coefficient, and $\eta(t)$ is a Gaussian white noise with zero mean and usual delta function correlations $\langle \eta(t)\eta(t') \rangle = \delta(t-t')$. We are interested in studying the probability distribution of the time integrated area:

\begin{equation}
\label{Avariable}
A = \int_0^T v^n(t) dt
\end{equation}
where $n$ is a positive integer. Some examples of this class of observables are schematically depicted in Figure \ref{fig:CTRW}. For $n=1$ the observable $A$ represents the total spatial displacement traveled by the particle up to time $T$, while for $n=2$, $A/T$ is the time averaged energy of the particle. Likewise, $A/T$ represents the time average of higher order moments, which in the long time limit converge to the corresponding ensemble averages. However, for finite though long times, fluctuation of $A$ are not trivial, and in particular similar observables are studied, for example, in the context of turbulent flows \cite{ApplicationsOU1, ApplicationsOU2}. As already observed in \cite{TouchetteInst, AlqahtaniInst, MeersonInst, NaftaliInst, HarrisInst, ChenInst}, for $n\leq 2$ fluctuations can be studied through a standard rate function in the framework of large deviation theory, while for $n>2$ at large time $T$ the standard large deviation approach does not hold. Nonetheless, in the small noise limit $\sigma \rightarrow 0$, or equivalently for $A$ sufficiently big, the description of rare events through a large deviation principle can be maintained, in the sense that it is possible to define a rate function, with an anomalous subexponential scaling:

\begin{equation}
 - \frac{\log P(A, T)}{T^{2/(2n-2)} } 
 \underset{A/T^{n/(2n-2)} \text{ fixed}}{\underset{A, T \rightarrow \infty}{\asymp}}
 \mathcal{J}\left(\frac{A}{T^{n/(2n-2)}}\right)
 \label{anomalous_rate}
 \end{equation}

with the system exhibiting a dynamical phase transition, since the generalized rate function $\mathcal{I}(x)$ displays a non-analytic singular point \cite{NaftaliInst}.

One of our goals is to study this behavior using the big jump principle, providing a simple relation between a single jump event and the asymptotic behavior of the anomalous rate function. For that aim, we map the Ornstein-Uhlenbeck process to a coupled CTRW process, where jump lengths and waiting times play a key role. We also use the Central Limit Theorem (CLT) and the renowned Einstein relation to find the typical fluctuation in this process.

In the following, we will focus on the case where $n$ is odd, where $\langle A(T) \rangle=0$. For $n$ even, as shown in Figure \ref{fig:CTRW}, the results can be obtained by subtracting from the observable $A$ the equilibrium average value $\langle A(T) \rangle = \langle v^n \rangle_{eq} T$. The approach for $n$ even will be discussed in Appendix \ref{AppendixC}, while in the main text we give the relevant final results.

\subsection{CTRW Mapping}

We now introduce the mapping of the process to a CTRW, the fundamental step to study the statistics of the observable $A$ at a given time $T$. The key idea is based on the concept of excursions \cite{Barkai2}. Our random process $v(t)$ is recurrent, and thus the velocity crosses the zero line $v=0$ many times in the time interval $(0, T)$ when $T$ is sufficiently large. In particular, if $v(t)$ is recurrent, also $v^n(t)$ is recurrent, and we can divide the observable $A$ into a sum of increments:

\begin{equation}
A = \int_0^{t_1} v^n(t')dt' + \int_{t_1}^{t_2} v^n(t')dt' + ... + \int_{t_i}^{t_{i+1}} v^n(t')dt' + ... + \int_{t_N}^{T} v^n(t')dt'    
\end{equation}
where $\{ t_1, t_2..., t_N \}$ are the times when the process crosses the zero line, i.e. $v^n(t_i)=0$, and in the $i$-th interval $(t_i, t_{i+1})$ the velocity is always above or below the zero line. Clearly, the process $v^n(t)$ in each of these intervals is a stochastic process that starts and ends in the chosen origin without any other crossing in between. Such a random curve is called an excursion \cite{Barkai2, MajumdarBrownianFunctional}. The value of the dynamical observable $A$ at time $T$ is now the sum of the areas under the excursions with their sign, each one integrated in the different crossing time intervals, so:

\begin{align}
\label{mappingodd}
T = \sum_{i=1}^N \tau_i + \tau^* \ \ \ &\text{with} \ \tau_i=t_{i+1}-t_i,  \ \tau^*= T - t_N \\
    A = \sum_{i=1}^N \mathcal{A}_i + \mathcal{A}^* \ \ \ &\text{with} \  \mathcal{A}_i=\int_{t_i}^{t_{i+1}} v^n(t')dt' , \  \mathcal{A}^*=\int_{t_N}^{T} v^n(t')dt' 
\end{align}

 We call the $i$-th time interval $\tau_i$ as the renewal time, and we call the area occurred in the $i$-th renewal time as the area under the excursion $\mathcal{A}_i$. 
This mapping allows us to analyze the continuous evolution in terms of a renewal process since every time the particle passes through the velocity origin, the process is renewed, and this is related to the Markovianity property of the Langevin dynamics. Therefore, the collection of couples of random variables $\{ \mathcal{A}_i, \tau_i \}_{i=1}^{N}$ univocally defines our CTRW.
As is well known in the literature \cite{CTRW1, CTRW2, CTRW3, CTRW4}, the CTRW extends the concept of a discrete random walker to a continuous time process in which the duration and the length of individual steps are drawn from a continuous probability distribution. Clearly, the number $N$ of zero crossings becomes a random variable, which depends on the measurement time, i.e. $N=N(T)$. 

For the CTRW, in the long time limit simple Einstein relations hold, see Appendix \ref{Appendix0} for details. This means that if the average time duration of an excursion $\langle \tau \rangle$ and its mean squared displacement $\langle \mathcal{A}^2 \rangle$ are finite, we can define an effective diffusion constant
$\langle A^2(T) \rangle = 2 D_n T$ and for $n$ odd we have:

\begin{equation}
\label{diffusion}
D_n = \frac{\langle\mathcal{A}^2\rangle}{2 \langle \tau \rangle},     
\end{equation}
In this context, the typical fluctuations of the PDF $P(A, T)$ at large $T$ will be a Gaussian as prescribed by the CLT, i.e. for $n$ odd:
\begin{equation}
\label{CLT_P}
P(A, T) \underset{T\rightarrow\infty}{\sim} \frac{1}{\sqrt{4\pi D_n T}} \exp\bigg\{ -\frac{A^2}{4 D_n T}\bigg\}
\end{equation}
The average duration $\langle \tau \rangle$ and the mean squared area $\langle \mathcal{A}^2 \rangle$ are simply defined by two ingredients: the PDF for the distribution of areas under the excursions $f(\mathcal{A})$, which play the role of jump-lengths, and the PDF for the distribution of waiting times $\psi(\tau)$.

Now, focusing on the rare events, the big jump principle has a very simple interpretation in this picture: the distribution of the rare events at large $A$ is completely characterized by the distribution of a single macroscopic area under one excursion that overpowers all other quantities by several orders of magnitude, and in a time $T$ in which an average number of jumps $\langle N(T) \rangle$ is performed, the PDF that the observable reaches the value $A$ at time $T$ is simply the product of the PDF that a single area of value $A$ occurs and the average number $\langle N(T) \rangle$ of attempts in which it can be performed \cite{BarkaiLaplaceErrors}, i.e:

\begin{equation}
\label{CTRW_BJ}
P(A, T) \underset{A \rightarrow \infty}{\sim} \langle N (T) \rangle f(A) \sim \frac{T}{\langle \tau \rangle} f(A)
\end{equation}

In the big jump principle approach \cite{BJ1} usually one has to treat in a separate way the processes where the big jump occurs in the last excursion described in equation \eqref{mappingodd} by area $\mathcal{A}^*$ and duration $\tau^*$ (backward recurrence time).  Here, however, as we will show, only the areas $\mathcal{A}$ are distributed sub-exponentially while the distribution of the durations $\tau$ exponentially decays to zero at large times. For this reason, when $T$ is large, i.e. $T\gg \langle \tau \rangle$ the probability that a big jump occurs in the last steps can be neglected, and the principle can be applied directly considering a set $\langle N(T) \rangle = T/\langle \tau \rangle$ of jumps where the jump length is drawn by a subexponential distribution $f(\mathcal{A})$. The big jump principle is valid provided that the PDF $f(\cal A)$ is subexponential. We will show that for $n>2$ this is precisely the case, in fact we will clarify why the anomalous scaling of the rate function \eqref{anomalous_rate} is deeply related to the fact that $f(\cal A)$ is subexponential and hence to the big jump principle.

To summarize, with the CTRW defined through the excursions and the times between zero crossings, we can fully characterize in a unified framework the typical fluctuations of the PDF $P(A,T)$ from the statistics of the moments of waiting times and excursion areas (\ref{diffusion},\ref{CLT_P}) and also its rare events, from the single big jump principle \eqref{CTRW_BJ}.

\begin{figure}
\includegraphics[scale = 0.6]{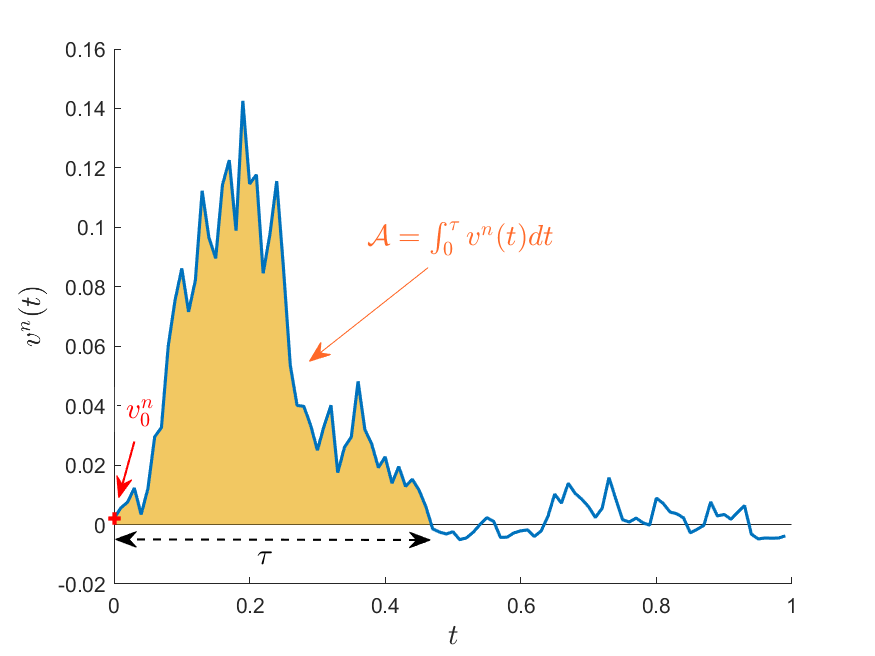}
\caption{A schematic figure showing a generic path of $v^n(t)$ starting at time $t=0$ at an initial velocity $v_0^n > 0$, indicated by the red arrow. The process travels on the positive region $(0, +\infty)$ until it first hits the origin after a time interval $\tau$, called first passage time, indicated by the double black arrow. The yellow area traveled until time $\tau$ corresponds to the area $\mathcal{A}$ under the first passage excursion, i.e. the first passage area;}
\label{fig:first_passage_area}
\end{figure}

We remark that in a continuous process the quantities we have introduced: $\langle \mathcal{A}^2\rangle$, $\langle \tau \rangle$ and $f(\mathcal{A})$ (for $\mathcal{A}>0$) are vanishing and $\langle N(T) \rangle$ is diverging as we will show in detail below. This is due to the fact that the paths are continuous and hence zero crossings are uncountable in a time interval $(0, T)$. However, excursions can be considered that do not start from $v=0$ but from $v_0>0$. In this case the same quantities are strictly positive and we expect that the ratios in equations (\ref{diffusion},\ref{CTRW_BJ}) remain finite when evaluated for $v_0>0$ and then the limit $v_0\to 0$ is taken into account.

In this context, we naturally consider
the PDF $\psi_{v_0}(\tau)$ and $f_{v_0}(\mathcal{A})$ for the durations and the areas under the excursions starting from $v_0>0$ and reaching $v=0$ for the first time (see Figure \ref{fig:first_passage_area} for a visual representation). Hence, these PDFs describe a first passage statistics for $\tau$ and the area $\mathcal{A}$ under the first passage path. These can be studied by introducing a backward Fokker-Planck equation with respect to the initial condition $v_0$ as discussed in the next section.

\subsection{First Passage Problems}

\subsubsection{First Passage Time Statistics $\psi_{v_0}(\tau)$}
\label{subsecfpt}

The PDF of first passage time $\psi_{v_0}(\tau)$ is a well known physical quantity studied with standard techniques \cite{Redner, Gardiner, VanKampen, Martingales_FPA}, in particular we want to compute the PDF $\psi_{v_0}(\tau)$ that a generic path of $v^n(t)$ starting from an initial velocity $v_0 \great 0$ travels the positive half-plane $(0, +\infty)$ until it hits for the first time the zero line in correspondence to the first passage time $\tau$. The first passage time to the absorbing point $v=0$ of the process $v^n(t)$ is the same as the first passage time of $v(t)$. Hence, we have:

\begin{equation}
\psi_{v_0}(\tau) = -\frac{\partial}{\partial \tau}\text{Prob}(t \geq \tau) \ \ \text{with $\tau= \inf \{t : v^n(t) \leq 0 \ | \  v^n(0)=v_0^n\}$}
\end{equation}
where here $\text{Prob}(t \geq \tau)$ is called the survival probability, and so, according to the first passage problems formalism \cite{Redner, Gardiner} the PDF $\psi_{v_0}(\tau)$ satisfies the following backward Fokker-Planck equation:

\begin{equation}
\label{pdfrenewaltimes}
\frac{\partial \psi_{v_0}(\tau)}{\partial \tau} = -\gamma v_0 \frac{\partial \psi_{v_0}(\tau)}{\partial v_0} + \frac{\sigma^2}{2}\frac{\partial^2 \psi_{v_0}(\tau)}{\partial v_0^2}
\end{equation}

Here, $\psi_{v_0}(\tau)$ satisfies the absorbing boundary condition on the zero line $\psi_{v_0 = 0}(\tau \neq 0) = 0$ and the reflective boundary condition at infinity $\partial_{v_0} \psi_{v_0 \rightarrow \infty}(\tau) = 0$. The solution of this backward equation is \cite{FPTimeOU1, FPTimeOU2, FPTimeOU3, FPTimeOU4, FPTimeOU0}:

\begin{equation}
\label{pdfrenewaltimessolution}
\psi_{v_0}(\tau) = \frac{\sqrt{\gamma}}{\sigma} v_0 \frac{\exp\bigg\{ -\frac{\gamma v_0^2}{\sigma^2}\frac{e^{-\gamma \tau}}{2\sinh\left(\gamma \tau\right)} + \frac{\gamma \tau}{2} \bigg\}}{\sqrt{2\pi \left(\sinh\left(\gamma\tau\right)\right)^3}}  
\end{equation}

As expected, at large $\tau$ an exponential decay can be observed, i.e. $\psi(\tau) \sim v_0 e^{-\gamma \tau}$.

\subsubsection{First Passage Area Under Excursions Statistics $f_{v_0}(\mathcal{A})$}

Analogously to what was done for the first passage times PDF, we can define a backward equation for the PDF of first passage areas under excursions $f_{v_0}(\mathcal{A})$. In Appendix \ref{AppendixA} we report the full derivation of the $f_{v_0}(\mathcal{A})$ equation from a path integral approach, firstly introduced in \cite{MajumdarBrownianFunctional, KearneyBrownianFunctional} to describe the first passage areas of Brownian particles. So, the backward Fokker-Planck equation for the first passage areas PDF is:

\begin{equation}
\label{pdffirstpassageareas}
v_0^n \frac{\partial f_{v_0}(\mathcal{A})}{\partial \mathcal{A}} - \gamma v_0 \frac{\partial f_{v_0}(\mathcal{A})}{\partial v_0} + \frac{\sigma^2}{2} \frac{\partial^2 f_{v_0}(\mathcal{A})}{\partial v_0^2} = 0 
\end{equation}

where $f_{v_0}(\mathcal{A})$ satisfies an absorbing boundary condition at the zero line $f_{v_0 = 0}(\mathcal{A} \neq 0) = 0$ and a reflective boundary condition at infinity $\partial_{v_0} f_{v_0 \rightarrow \infty}(\mathcal{A}) = 0$. As mentioned, using the big jump principle \eqref{CTRW_BJ}, the solution of this equation provides interesting physical insights about the asymptotic behavior of the PDF of the global process $P(A, T)$ for different values of $A$. 

\subsection{Main Results}
\label{sub_mainresults}

We summarize our main results. We first study the typical fluctuations of the PDF $P(A, T)$ in Section \ref{Section3} in the long time limit, i.e. we characterize the bulk distribution of the process which will be a Gaussian in accordance with the CLT. We calculate the diffusion coefficient $D_n$ using the CTRW mapping described above, in particular we use Einstein's relation \eqref{diffusion}, and from the calculation of the moments of the first passage PDFs $\langle \tau \rangle$, $\langle \mathcal{A} \rangle$ and the variance of the area under the first passage excursion $\langle \mathcal{A}^2 \rangle$ we derive the "diffusion constant", finding two different results for odd and even $n$ \footnote{For even and odd $n$ we use $\langle A ^2 \rangle - \langle A \rangle ^2 \sim 2 D_n T$ where $\langle A \rangle = 0$ for odd $n$.}:

\begin{equation}
    D_n =
    \begin{cases}
    \frac{\sigma^{2n}}{2^n \gamma^{n+1}} \sum_{j=0}^{\frac{n-1}{2}} \frac{(n-1)!!}{2j!!} \frac{(n+2j)!!}{2j+1} & \text{for $n$ odd} \\ 
    \frac{\sigma^{2n}}{2^n \gamma^{n+1}} \sum_{j=0}^{\frac{n}{2}-1} \frac{(n-1)!!}{(2j+1)!!} \frac{\left((n+2j+1)!! - (n-1)!!(2j+1)!! \right)}{2j+2}  & \text{for $n$ even}
    \end{cases}
    \label{Dn}
\end{equation}
where the mathematical symbol $m!! = \prod_{i=0}^{m/2-1}(m-2i)=m(m-2)(m-4)...$ stands for the double factorial.

The results for $D_n$ obtained from our CTRW recovers the expression computed in \cite{NaftaliInst} through a large deviation technique based on a quantum mapping of the stochastic process in an effective Schr$\ddot{o}$dinger equation, the so called D$\ddot{o}$nsker-Varadhan formalism. 
As shown in \cite{NaftaliInst}, the Ornstein-Uhlenbeck process for the observable $A$ can be viewed as a quantum one dimensional harmonic oscillator perturbed by the action of an external non-confining potential $v^n$ with $n \great 2$, and the diffusion coefficent coincides with the second-order perturbative expansion of the energy states. This perturbative approach is valid as long as typical fluctuations of the observable are explored, and becomes ineffective when rare events need to be studied, as the non-confining potential can no longer be considered as a small perturbation around the harmonic solution. We point out that the result for $D_n$ obtained in \cite{NaftaliInst} is valid only for $n>2$, The result for $D_n$ obtained in \cite{NaftaliInst} is valid only for $n>2$, since it comes out of a perturbative approach. In contrast, our formula for the diffusive constant holds for any value of $n$. In particular, it is true for $n=1, 2$.
In Appendix \ref{AppendixB} we show that the analytical result for the diffusivity can also be obtained from the Green-Kubo formula \cite{Barkai_GreenKubo1, Barkai_GreenKubo2, GreenKubo3}.

Then, in Section \ref{Section4} we explore the solutions of the first passage area PDF $f_{v_0}(\mathcal{A})$ through an approach based on a formalism of large deviations \cite{BarkaiStretchedExp, BarkaiStretchedExp2}, where we are able to define the corresponding anomalous rate function $\cal{I}(.)$:

\begin{equation}
\label{pdf_rate_calA}
f_{v_0}(\mathcal{A}) \underset{v_0/\mathcal{A}^{1/n} \text{ fixed}}{\underset{\mathcal{A} \rightarrow \infty}{\asymp}} v_0 \exp \bigg \{ - \frac{\mathcal{A}^{2/n}}{\sigma^2}  \mathcal{I}\left(\frac{v_0}{\mathcal{A}^{1/n}}\right) \bigg \}
\end{equation}

This is technically the main part of our work. In the analysis of the solutions of the first passage PDF $f_{v_0}(\mathcal{A})$ we discover two different asymptotic regimes: a Brownian diffusion regime \cite{MajumdarFPA} when $\mathcal{A} \rightarrow 0$, or equivalently when we are in a strong noise regime $\sigma \rightarrow \infty$, and an anomalous subexponential regime when $\mathcal{A} \rightarrow \infty$, i.e. in the limit of weak noise $\sigma \rightarrow 0$. The final result is:

\begin{equation}
\label{pdfAgeneral}
f_{v_0}(\mathcal{A}) \underset{v_0 \rightarrow 0}{\asymp}
\begin{cases}
v_0 \mathcal{A}^{-\frac{n+3}{n+2}} \exp \bigg\{ -\frac{2}{(n+2)^2 \sigma^2} \frac{v_0^{n+2}}{\mathcal{A}}\bigg\} & \text{for $\mathcal{A} \rightarrow 0$} \\
v_0 \exp \bigg\{ -\frac{\gamma^{\frac{n+2}{n}}}{\sigma^2} c_n \mathcal{A}^{\frac{2}{n}} \bigg\} & \text{for $\mathcal{A} \rightarrow \infty$} 
\end{cases}
\end{equation}

In particular these results for $\mathcal{A} \rightarrow 0$ are valid for any $n$, though the main focus of our work is when $n>2$. The numerical constant $c_n$ is reported in Equation \eqref{coeffcn}. We also show for the first time that the rate function describing the area under the first passage excursion exhibits a critical point, that corresponds to the noiseless solution of the process and separates these two asymptotic regimes. This will allow us to obtain interesting physical insights into the behaviour of the global process, by connecting our new results for the first passage PDF $f_{v_0}(\mathcal{A})$ with the dynamical phase transition of the global PDF $P(A, T)$ first observed in \cite{NaftaliInst}.

Finally, in Section \ref{Section5} we will focus on the study of rare events of the initial PDF $P(A,T)$ through our CTRW mapping and by direct use of the big jump principle. In fact, since we may interpret in the large $A$ limit our initial process as a CTRW whose distribution of the jump lenghts, i.e. the areas under excursions, is sub-exponential \cite{BarkaiLaplaceErrors}, the single big jump principle holds and prescribes a simple statement: the probability that the time-integrated variable \eqref{Avariable}, with $n>2$, reach the value $A$ in a time $T$ is equal to the product of the probability that a single big jump of area $A$ occurs times the number of trials that occurred in time $T$. Using equations (\ref{CTRW_BJ},\ref{pdfAgeneral}) this lead us to the following result in the long time limit:

\begin{equation}
P(A, T) \underset{A \rightarrow \infty}{\asymp} T \exp\bigg \{ - \frac{\gamma^{\frac{n+2}{n}}}{\sigma^2} c_n A^{\frac{2}{n}}  \bigg \}
\end{equation}
where $c_n$ is \eqref{coeffcn}, and the result obtained coincides with what has already been found in the works of \cite{TouchetteInst, MeersonInst, NaftaliInst}, and allows us to demonstrate a formal analogy between the single big jump approach and the instantonic approach. In addition, our result for $P(A, T)$ contains a linear subexponential factor in $T$, related to the average number of renewals performed in the total time and to the law of large numbers. The asymptotic expression of $P(A, T)$ may also display a subleading prefactor which depends on $A$. Our numerical results show that such a term is indeed determined by the presence of an unknown prefactor  even in Eq. \eqref{pdfAgeneral} at large  ${\cal A}$; i.e. we observe that the big jump formula \eqref{CTRW_BJ} holds for values of $A$ where Eq. \eqref{pdfAgeneral} for $f_{v_=}(\mathcal{A})$ is preasymptotic and the presence of a subleading prefactor cannot be neglected. Interestingly, this subleading pre-factors should be related to the Gaussian correction term of the small noise limit in the instantonic approach, as reported in \cite{TouchetteInst_Corr}. Discussions on this link, possible generalisations on a broader class of stochastic processes, future perspectives and open questions can be found at the end in Section \ref{Section6}.

\section{Typical Fluctuations}
\label{Section3}

In this section we derive the second moment of the first passage areas under the excursions $\langle \mathcal{A}^2 \rangle$ and the first moment of the first passage times $\langle \tau \rangle$ by using the relevant backward equations (\ref{pdfrenewaltimes},\ref{pdffirstpassageareas}) for the first passage time PDF $\psi_{v_0}(\tau)$ and the first passage excursion area 
$f_{v_0}(\mathcal{A})$. From these results, one can obtain the value of the diffusion coefficient $D_n$ from equation \eqref{diffusion} and the expression for the typical Gaussian fluctuation of the PDF in \eqref{CLT_P}.

We start by calculating the first moment of the renewal times $\langle \tau \rangle$. A simple way to derive the mean value is to return to the problem of the first passage time PDF $\psi_{v_0}(\tau)$ and define the mean first passage time as $\langle \tau \rangle(v_0) = \int_{0}^{+\infty} d\tau \ \tau \  \psi_{v_0}(\tau)$. Note that this expectation value, being related to a first passage problem, depends on the value of the initial condition $v_0$. Now, since we are interested in calculating the statistics of the mean renewal times around the zero line, we can derive the desired mean value simply by taking the limit for $v_0 \rightarrow 0$ of the mean first passage time. At this point, deriving a backward equation for $\langle \tau \rangle (v_0)$ is simple, just take the equation \eqref{pdfrenewaltimes}, multiply it by $\tau$ and integrate it with respect to time. Then, we have that:

\begin{equation}
\label{equationmfpt}
-\gamma v_0 \frac{\partial}{\partial v_0} \langle \tau \rangle (v_0) + \frac{\sigma^2}{2} \frac{\partial^2}{\partial v_0^2} \langle \tau \rangle (v_0) = -1
\end{equation}

The mean first passage time $ \langle \tau \rangle(v_0)$ satisfies an absorbing boundary condition at the zero line $\langle \tau \rangle (v_0=0)=0$ and a reflective boundary condition at infinity $\partial_{v_0} \langle \tau \rangle (v_0 \rightarrow \infty) =0$. This equation can be easily solved analytically via the variation of constants method, and the asymptotic approximated solution of the mean first passage time $\langle \tau \rangle (v_0)$ for $v_0 \rightarrow 0$, which corresponds to the mean renewal time, is (see Appendix \ref{AppendixC} for more details and for the full solution of the mean renewal time):

\begin{equation}
\label{meanfpt}
\langle \tau \rangle (v_0)  \underset{v_0 \rightarrow 0}{\sim} \frac{1}{\sigma} \sqrt{\frac{\pi}{\gamma}} v_0
\end{equation}

As expected, the leading behavior of the mean renewal time is linear in $v_0$, while as $\sigma \rightarrow 0$ the time to reach the absorbing boundary condition diverges, since then the relaxation process is deterministic and exponential, namely $v(t)=v_0 e^{-\gamma t}$, and the particle strictly reaches the zero line only after an infinite time.

Now we can move on to calculate the mean area under the excursion $ \langle \mathcal{A}\rangle$, and to do it we start from first passage excursion area PDF $f_{v_0}(\mathcal{A})$ backward equation \eqref{pdffirstpassageareas}. Similarly to what we did before, we can move to the equation for its first moment by defining the mean first passage excursion as $\langle \mathcal{A} \rangle (v_0) = 2 \int_{0}^{+\infty} d\mathcal{A} \ \mathcal{A} f_{v_0}(\mathcal{A})$ , where the factor 2 takes into account the fact that first passage excursions can have positive or negative area, depending on the sign of the initial condition $v_0$. At this point, integrating in $\mathcal{A}$ the backward equation for the PDF we get:

\begin{equation}
\label{equationmfpexcursion}
-\gamma v_0 \frac{\partial}{\partial v_0} \langle \mathcal{A} \rangle (v_0) + \frac{\sigma^2}{2} \frac{\partial^2}{\partial v_0^2} \langle \mathcal{A} \rangle (v_0) = 2v_0^n
\end{equation}

With absorbing boundary condition at the zero line $\langle \mathcal{A} \rangle (v_0=0)=0$ and reflective boundary condition at infinity $\partial_{v_0} \langle \mathcal{A} \rangle (v_0\rightarrow \infty) =0$. Also, this equation can be solved explicitly by the variation of constants method, and we can distinguish two general expressions, depending on whether $n$ is odd or even. Again, all details of the calculations are explained in Appendix \ref{AppendixC}, and here we report the final result:

\begin{equation}
\label{meanexcursion}
\langle \mathcal{A} \rangle (v_0) = 
\begin{cases}
2 \frac{\sigma^{n-1}}{2^{\frac{n-1}{2}}\gamma^{\frac{n+1}{2}}} \sum_{j=0}^{\frac{n-1}{2}} \frac{(n-1)!!}{(2j)!!} \frac{2^j \gamma^j}{\sigma^{2j}} \frac{v_0^{2j+1}}{2j+1} & \text{for $n$ odd}\\
2 \frac{\sigma^{n-1}}{2^{\frac{n-1}{2}}\gamma^{\frac{n+1}{2}}} \sum_{j=0}^{\frac{n}{2}-1} \frac{(n-1)!!}{(2j+1)!!} \frac{2^{j+\frac{1}{2}} \gamma^{j+\frac{1}{2}}}{\sigma^{2j+1}} \frac{v_0^{2j+2}}{2j+2} & \text{for $n$ even}
\end{cases}
\end{equation}

Note that this result for the first moment of $\mathcal{A}$ is general and valid for any initial condition $v_0$. 

At this point we can calculate the mean squared first passage excursion $\langle \mathcal{A}^2 \rangle$ of the process, and again, we note that the second moment associated with the PDF $f_{v_0}(\mathcal{A})$ is defined as $\langle \mathcal{A}^2 \rangle= \langle \mathcal{A}^2 \rangle(v_0) = 2 \int_0^{+\infty} d\mathcal{A} \ \mathcal{A}^2 P(\mathcal{A}, v_0)$, where the factor 2 is due to the fact that fluctuations in first passage excursions can be positive or negative. Also for the second moment it is possible to write a backward equation, which will depend by the first moment $\langle \mathcal{A} \rangle$, since there is a recursive relation between the equation of the $n$-th moment $\langle \mathcal{A}^n \rangle$ with the $(n-1)$-th moment $\langle \mathcal{A}^{n-1} \rangle$ of the first passage area, as reported in \cite{Gardiner}. So, we find:

\begin{equation}
\label{equationmeansquaredexc}
-\gamma v_0 \frac{\partial}{\partial v_0} \langle \mathcal{A}^2 \rangle(v_0) + \frac{\sigma^2}{2} \frac{\partial^2}{\partial v_0^2} \langle \mathcal{A}^2 \rangle (v_0) = 2v_0^n\langle \mathcal{A} \rangle (v_0)
\end{equation}

\begin{figure}
\includegraphics[scale = 0.6]{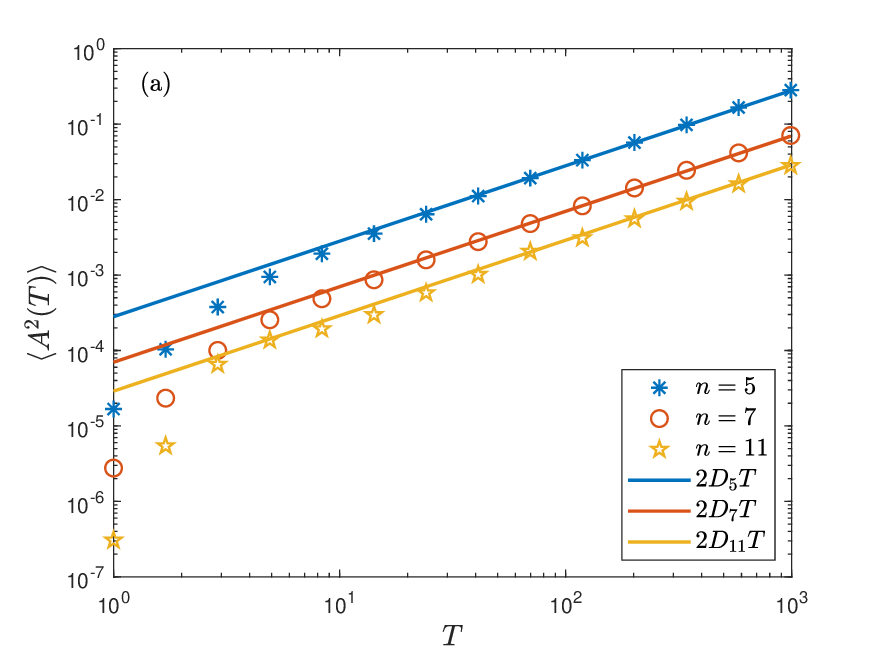}
\includegraphics[scale = 0.6]{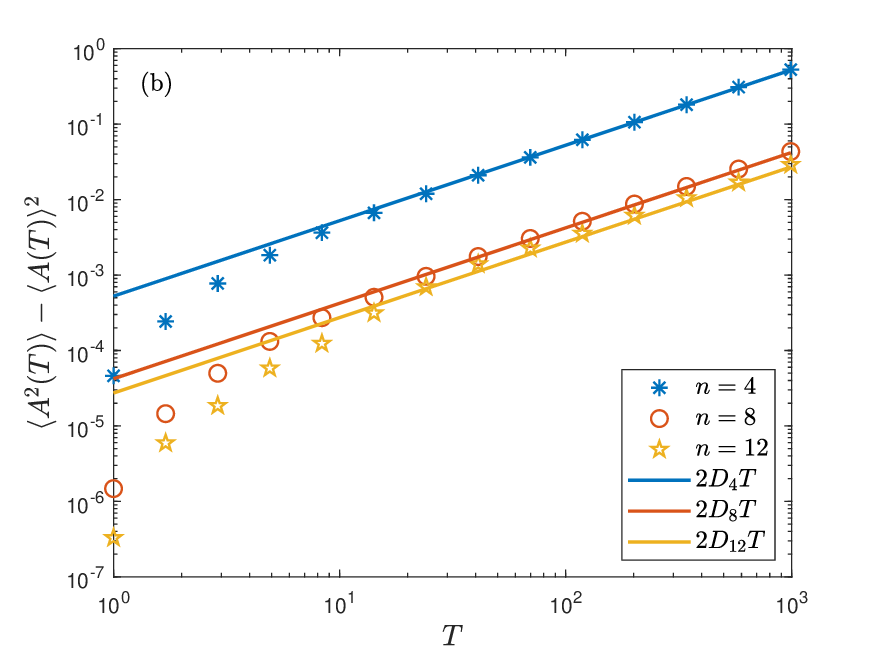}
\caption{The variance $\text{Var}(A)$ plotted as a function of time $T$ for different values of $n$. Figure (a) is the case of odd $n$, and shows numerical simulations of the second moment compared with analytical prediction $\text{Var}(A)=\langle A^2(T) \rangle =2 D_n T$. Figure (b) is the case of even $n$, and shows numerical simulations of the second moment shifted by the squared expectation value of the observable compared with the analytical prediction $\text{Var}(A)=\langle A^2(T) \rangle - \langle A(T) \rangle^2 =2 D_n T$. The values of the diffusion coefficients $D_n$ were computed using the formula \eqref{diffusion} and in all simulations the parameters were set to $\gamma=1$, $\sigma=1$;}
\label{fig:Diffusion}
\end{figure}
where the mean square first passage excursion satisfies the absorbing boundary condition at zero $\langle \mathcal{A}^2 \rangle (v_0=0) = 0$ and the reflective boundary condition at infinity $\partial_{v_0} \langle \mathcal{A}^2 \rangle(v_0\rightarrow \infty) = 0$. This equation has a general solution that is explicitly dependent on the value of the first moment of the first passage area:

\begin{equation}
\langle \mathcal{A}^2 \rangle (v_0) = 2 \int_0^{v_0} dy \ e^{-\frac{\gamma y^2}{\sigma^2}} \int_{y}^{+\infty} dv \ v^n  e^{-\frac{\gamma v^2}{\sigma^2}}\langle \mathcal{A} \rangle(v)
\end{equation}

Now, in the limit of $v_0 \rightarrow 0$ this integral can be approximated, since the asymptotics of the Gaussian integral $\int_0^{v_0} dy e^{-\frac{\gamma y^2}{\sigma^2}} \sim v_0$ can be linearized, and the extremes of the inner integral now go from zero to infinity, and thus the asymptotic expression of $\langle \mathcal{A}^2 \rangle(v_0)$ for $v_0$ small is derived:

\begin{equation}
\langle \mathcal{A}^2 \rangle (v_0)  \underset{v_0 \rightarrow 0}{\sim}2 v_0 \int_{0}^{+\infty} dv \ v^n e^{-\frac{\gamma v^2}{\sigma^2}} \langle \mathcal{A} \rangle(v)
\end{equation}

By substituting the analytical expression of $\langle \mathcal{A} \rangle (v_0)$ found in \eqref{meanexcursion} it is possible to calculate explicitly this integral. The solution has two different forms, depending on whether $n$ is even or odd. Again, all the details are left in the Appendix \ref{AppendixC}. We finally have these solutions:

\begin{equation}
\langle \mathcal{A}^2 \rangle (v_0) \underset{v_0 \rightarrow 0}{\sim}
\begin{cases}
    2 v_0 \frac{\sigma^{2n}}{2^n \gamma^{n+1}} \sum_{j=0}^{\frac{n-1}{2}} \frac{(n-1)!!}{2j!!} \frac{(n+2j)!!}{2j+1} \sqrt{\frac{\pi}{\gamma}} \sigma & \text{for $n$ odd} \\ 
    2 v_0 \frac{\sigma^{2n}}{2^n \gamma^{n+1}} \sum_{j=0}^{\frac{n}{2}-1} \frac{(n-1)!!}{(2j+1)!!} \frac{\left((n+2j+1)!! - (n-1)!!(2j+1)!! \right)}{2j+2} \sqrt{\frac{\pi}{\gamma}} \sigma & \text{for $n$ even}
\end{cases}
\label{meanA2}
\end{equation}

Now, we can plug equations (\ref{meanfpt},\ref{meanA2}) for $\langle \mathcal \tau \rangle (v_0)$ and $\langle \mathcal{A}^2 \rangle (v_0)$ obtained in the limit of small $v_0$ into equation (\ref{diffusion}) and in the limit $v_0\to 0$ we get the analytic expression for the diffusivity $D_n$ \eqref{Dn}.
The result perfectly matches with numerical simulations as shown in Figure \ref{fig:Diffusion}.

\section{Area Under the First Passage Excursion}
\label{Section4}

\noindent In this section we study the probability distribution of the area under the first passage excursion $\mathcal{A}=\int_0^{\tau} v^n(t)dt$, where $\tau$ is a first passage time. We set an arbitrary initial velocity $v_0 > 0$. Since the sign of $v(t)$ does not change in the interval $[0,\tau]$ we have $f_{v_0}(\mathcal{A})=0$ for $v_0>0$ and $\mathcal{A}<0$, moreover the parity symmetries $f_{-v_0}(-\mathcal{A})=f_{v_0}(\mathcal{A})$ for $n$ odd and $f_{-v_0}(\mathcal{A})=f_{v_0}(\mathcal{A})$ for $n$ even and the absorbing condition $f_{v_0=0}(\mathcal{A})=\delta(\mathcal{A})$ holds. 

Let us consider the backward Fokker-Planck equation \eqref{pdffirstpassageareas}. A scaling approach, which provides important information about the physics of the system, is based on a large deviations formalism, where the PDF $f_{v_0}(\mathcal{A})$ takes the form of the exponential of a generalized rate function: 
\begin{equation}
\label{ansatz}
f_{v_0}(\mathcal{A}) \underset{v_0/\mathcal{A}^{\beta} \text{ fixed}}{\underset{\mathcal{A} \rightarrow \infty}{\asymp}} v_0 \exp \bigg \{ - \frac{\mathcal{A}^{\delta}}{\sigma^2}  \mathcal{I}\left(\frac{v_0}{\mathcal{A}^{\beta}}\right) \bigg \}. 
\end{equation}
\noindent where the linear term $v_0$ in the pre-factor takes into account of the absorbing boundary condition. We highlight the dependence on noise amplitude $\sigma$. The two exponents  $\beta$ and $\delta$ characterize our ansatz. By plugging the rate function ansatz \eqref{ansatz} into the backward equation \eqref{pdffirstpassageareas} and by keeping the leading terms in the limit of $\mathcal{A} \rightarrow \infty$, or equivalently in the small noise limit $\sigma \rightarrow 0$, we obtain (see Appendix \ref{AppendixD} for details):

\begin{equation}
\frac{ \omega(\mathcal{I}'(\omega))^2}{2 \mathcal{A}^{-\delta + \beta}} +\frac{\gamma \omega^2 \mathcal{I}'(\omega)}{\mathcal{A}^{-\beta}} + \frac{\delta \omega^{n+1} \mathcal{I}(\omega) + \beta \omega^{n+2}\mathcal{I}'(\omega)}{\mathcal{A}^{-\beta n - \beta + 1}} = 0.
\end{equation}

\noindent with $\omega={v_0}/{\mathcal{A}^{\beta}}$. Now, the three terms should display the same dependence on $\mathcal{A}$, therefore:
\begin{equation}
\label{scalingexpo}
\beta = \frac{1}{n} \ \ \text{and} \ \ \delta= 2 \beta = \frac{2}{n} 
\end{equation}

\noindent i.e. the result in Equation \eqref{pdf_rate_calA}. Moreover, the rate function $\mathcal{I}(\omega)$ is governed by the following non-linear differential equation:

\begin{equation}
\label{ratefunctionequation}
\frac{1}{2} (\mathcal{I}'(\omega))^2 + \mathcal{I}'(\omega)\left( \gamma \omega - \frac{\omega^{n+1}}{n} \right) +\frac{2}{n}\omega^n \mathcal{I}(\omega) = 0.
\end{equation}

\begin{figure}
\includegraphics[scale = 0.6]{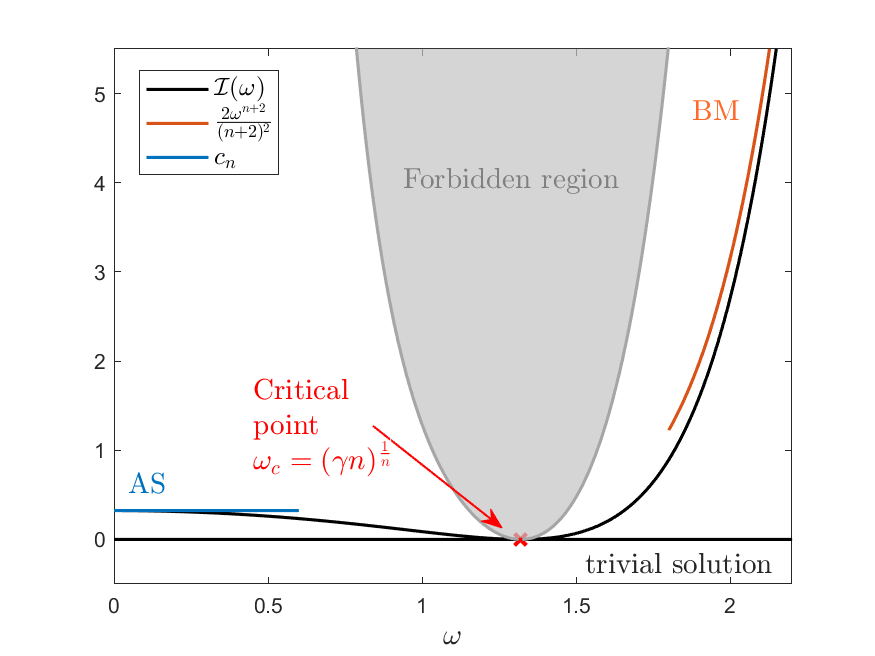}
\caption{Schematic plot of the asymptotic solutions of the rate function equation $\mathcal{I}(\omega)$ \eqref{ratefunctionequation} versus the scaling variable $\omega=v_0/\mathcal{A}^{1/n}$, where in this case $n=7$. The rate function has in general two asymptotic solutions, represented by the continuous black lines. One is the trivial solution $I(\omega)=0$, while the other is the non-trivial solution of our interest. We identify with the grey area the forbidden region where the discriminant of \eqref{ratefunctionequation} is negative, while the critical point where the two stationary solutions match is indicated with the red arrow. The non-zero solution has two interesting asymptotic trends: in the limit of $\omega \rightarrow \infty$ (i.e. $\mathcal{A} \rightarrow 0$) we find the diffusive scaling typical of Brownian motion \eqref{pdfAgeneral}, denoted as BM, while in the limit of $\omega \rightarrow 0$ we find the anomalous sub-exponential scaling \eqref{coeffcn}, denoted as AS;}
\label{fig:ratefunction_analytical}
\end{figure}

\noindent
Equation \eqref{ratefunctionequation} has an infinite number of solutions. In particular, since the equation is quadratic in $\mathcal{I}'(\omega)$, there exist two solutions for each value of $\omega$ and $\mathcal{I}(\omega)$.
The discriminant of equation \eqref{ratefunctionequation} is $n^{-1}(\gamma \omega n -\omega^{n+1})^2 - 4 \omega^n \mathcal{I}(\omega)$, and discussing its sign we obtain that for $I(\omega)>n(4 \omega^n)^{-1}(n \gamma \omega -\omega^{n+1})^2$ there are no real solutions. 
On the boundary of such forbidden region typically two real solutions of \eqref{ratefunctionequation} merge with the same slope (zero discriminant) but they cannot be further extended where the sign of the discriminant is negative. As illustrated in Figure \ref{fig:ratefunction_analytical}, for $n>2$ a  different behavior is observed only at a critical point where $\omega=\omega_c=(\gamma n)^{{1}/{n}}$ and $\mathcal{I}(\omega_c)=0$; here two real solutions (defined for all $\omega>0$) are tangent to the forbidden region. One is the trivial solution $\mathcal{I}(\omega)=0$ and the other can be evaluated by integrating numerically equation \eqref{ratefunctionequation}.

Let us show that only this second solution has a physical relevance for our problem. In the limit of vanishing noise $\sigma\to 0$, the Ornstein-Uhlenbeck process \eqref{OUequation} is determined by the exponential decay $v(t)=v_0 e^{-\gamma t}$, without any stochastic fluctuation. So, the first passage area in this limit becomes $\mathcal{A}=\int_0^{\infty}v_0^n e^{-n\gamma t}dt = v_0^n/(\gamma n)$ and we get $\omega=\omega_c$ (time integration is until infinity since our process is deterministic). 
Therefore, according to the large deviation ansatz in Equation \eqref{ansatz}, a saddle point approach can be applied when $\sigma \to 0$ and the physical solution $\mathcal{I}(\omega)$ of Equation \eqref{ratefunctionequation} shows an absolute minimum at $\omega=\omega_c$. 
The analysis in Appendix \ref{AppendixD} shows that for $\omega>0$ we can have $\mathcal{I}'(\omega)=0$ and $\mathcal{I}''(\omega)>0$ only at the critical point. The physical solution hence can be expanded at $\omega_c$ as $\mathcal{I}(\omega)\approx \gamma (n-1) (\omega-\omega_c)^2$ so that we obtain the Gaussian approximation for $f_{v_0}(\mathcal{A})$ which describes the small fluctuations around the noiseless solution. Since we are interested in the case of large $\mathcal{A}$ (i.e. $\omega \to 0$) we observe that in the physical solution $\mathcal{I}(\omega)$ tends to a constant at small $\omega$. So we can write
$\mathcal{I}(0)=c_n \gamma^{\frac{n+2}{n}}$ and reproduce the behavior in \eqref{pdfAgeneral} for large $\mathcal{A}$. 

We also observe that with a suitable change of variables, equation \eqref{ratefunctionequation} can be mapped into the differential equation for the rate function in \cite{BarkaiStretchedExp, BarkaiStretchedExp2}, which describes the approach to the equilibrium for a Brownian particle in a weakly binding potential, such that the Boltzmann equilibrium density is stretched exponential (see Appendix \ref{AppendixD} for details). Interestingly,  in 
\cite{BarkaiStretchedExp, BarkaiStretchedExp2} it is shown analytically that the physical solution tends to a constant at small $\omega$ and in particular they present an analytic expression for $\mathcal{I}(0)$ observed in our numerical solution:
\begin{equation}
\label{coeffcn}
\mathcal{I}(0) = \gamma^{\frac{n+2}{n}} \frac{n}{4} \left[ \frac{2\sqrt{\pi}}{n-2} \frac{\Gamma \left( \frac{n}{n-2} \right)}{ \Gamma \left(  \frac{3n-2}{2n-4} \right)} \right] \equiv \gamma^{\frac{n+2}{n}} c_n.
\end{equation}
and this gives us the constant $c_n$ of equation \eqref{pdfAgeneral} for $\mathcal{A} \rightarrow \infty$, which is in agreement with the results previously obtained in \cite{TouchetteInst, MeersonInst, NaftaliInst}. Here, unlike in the case considered in \cite{BarkaiStretchedExp, BarkaiStretchedExp2}, at the critical point there is no non-analyticality, given by the switch between the two solutions of the rate function equation.

Now we finally study the behavior of the solution at large $\omega$, i.e. at small $\mathcal{A}$. By introducing the asymptotic ansatz:
\begin{equation}
\mathcal{I}(\omega)\sim C \omega^{\theta} 
\end{equation}
into the equation \eqref{ratefunctionequation} and considering only the leading terms in the limit of $\omega \rightarrow \infty$ we can fix the parameters:
\begin{equation}
\frac{C\theta^2}{2}\omega^{2\theta-2} + \frac{2-\theta}{n}\omega^{n+\theta} = 0 \ \rightarrow \ \theta=n+2, \ C=\frac{2}{(n+2)^2}.
\end{equation}
So we recover the exponential behavior at small $\mathcal{A}$ in equation \eqref{pdfAgeneral}. We remark that this coincides with the cut-off in the solution of the first passage area under the excursion PDF of a Brownian motion, as reported in \cite{MajumdarFPA}. This is not surprising since for large $\omega$, i.e. small $\mathcal{A}$, the effect of the deterministic linear force $-\gamma v$ in the starting Langevin equation \eqref{OUequation} should be negligible and indeed the result is independent of $\gamma$. In this perspective we expect that, at small enough $\mathcal{A}$, also the prefactor should be described by the behavior of the Brownian motion in \cite{MajumdarFPA}, obtaining the power-law behavior $\mathcal{A}^{-\frac{n+3}{n+2}}$ in \eqref{pdfAgeneral}.

\begin{figure}
\includegraphics[scale = 0.6]{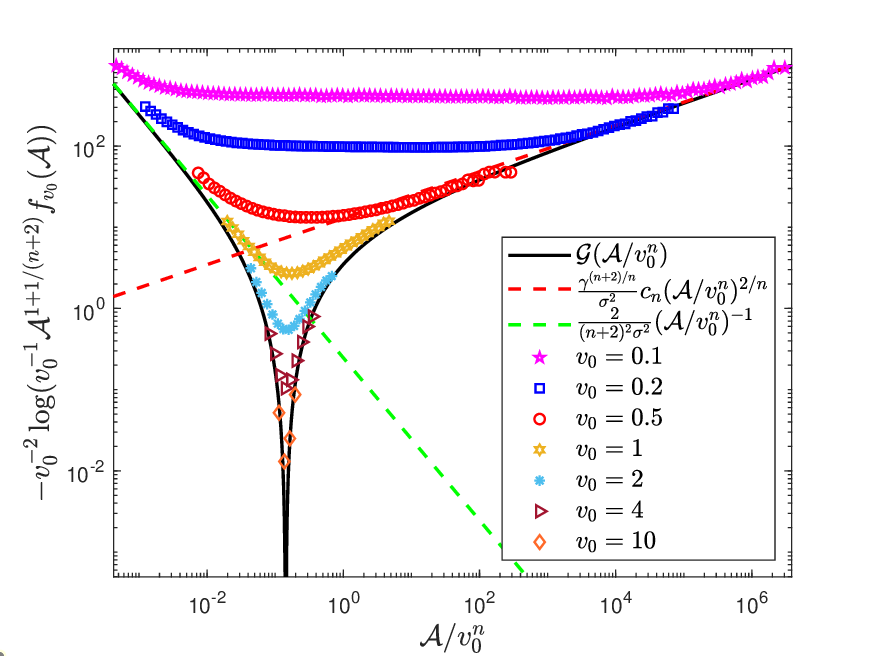}
\caption{Plot of the rate function of the area under the first passage excursion for the Ornstein-Uhlenbeck process with $n=7$ in logarithmic scale. The solid black line represents the analytical expression of the asymptotic non trivial solution of the rate function integrated numerically from equation \eqref{ratefunctionequation}, and here it is plotted in terms of the rescaled function $\mathcal{G}(\mathcal{A}/v_0^n)$, defined as $\mathcal{G}(z)=  z^{\frac{2}{n}} \mathcal{I}\left( z^{-1/n}\right)$. $\mathcal{G}(z)$ display a smooth minimum at $\mathcal{A}/v_0^n=\omega_c^n$ which is magnified by the logarithmic scale. The dashed red line represents the sub-exponential anomalous behavior of the rate function in the limit of $\mathcal{A}/v_0^n \rightarrow \infty$ and the dashed green line represents the diffusive Brownian motion-like behavior of the rate function in the limit of $\mathcal{A}/v_0^n \rightarrow 0$, in accordance with the analytical results \eqref{pdfAgeneral}. The dot curves show some simulations for the statistics of first passage areas $\mathcal{A}$ at different values of $v_0$. All these simulations were carried out by setting $\gamma=1$ and $\sigma=0.31682$.}
\label{fig:ratefunction_full}
\end{figure}  

In the numerical simulations of the Ornstein-Uhlenbeck process, one evaluates the distribution of $\mathcal{A}$
for a fixed initial condition $v_0$. In this perspective 
the rate function ansatz in equation \eqref{ansatz} is naturally written as 
\begin{equation}
\label{ansatz2}
f_{v_0}(\mathcal{A}) \asymp v_0 \mathcal{A}^{-\frac{n+3}{n+2}} \exp \bigg \{ - \frac{{v_0}^2}{\sigma^2}  \mathcal{G}\left(\frac{\mathcal{A}}{v_0^{n}}\right) \bigg \} 
\end{equation}
where $\delta$ and $\beta$ have been fixed by equation \eqref{scalingexpo}, and the new scaling function is given by $\mathcal{G}(z)=  z^{\frac{2}{n}} \mathcal{I}\left( z^{-1/n}\right)$. The prefactors $v_0$
and $\mathcal{A}^{-\frac{n+3}{n+2}}$ take into account of the boundary condition at $v_0=0$ and of the expected behavior at small $\mathcal{A}$ respectively. Clearly in the regime where we expect that the rate function ansatz holds (large $\mathcal{A}$ and $v_0$ with  fixed $\mathcal{A}/v_0^n$),  one can neglect the prefactors.

The results of the numerical simulations are shown in Figure \ref{fig:ratefunction_full}. For large $v_0$ (and then also large $\mathcal{A}$) we find that numerical data are consistent with the analytic solution for the rate function, however, as expected in this regime fluctuations are very small and $f_{v_0}(\mathcal{A})$ is very peaked around the solution without noise at $\mathcal{A}/v_0^n=\omega_c^{-n}$; therefore, in the numerical simulations the statistic of rare events at large and small $\mathcal{A}$ cannot be measured. On the other hand, for small $v_0$, the equation \eqref{ratefunctionequation} does not describe the distribution $f_{v_0}(\mathcal{A})$, in particular differences emerge especially when $\mathcal{A}/v_0^n$ is close to the noiseless minimum $\omega_c^{-n}$. For small $v_0$ indeed, noise can never be considered small and $f_{v_0}(\mathcal{A})$ cannot be described by small fluctuations around the noiseless solution. Remarkably, numerical simulations show that for small $v_0$ the analytical solution of equation \eqref{ratefunctionequation} matches the distribution $f_{v_0}(\mathcal{A})$ in both large and small $\mathcal{A}$. At small $\mathcal{A}$ we expect that the Brownian motion describes the system well independently of $v_0$. On the other hand, when $\mathcal{A}$ is large, the dynamics is expected to be independent of the initial conditions, and therefore the result obtained for large $v_0$ holds also for small $v_0$.

\section{Single Big Jump Approach}
\label{Section5}

Now we are finally ready to characterise the rare events of the dynamical observable $A$ integrated until a time $T$ by deriving the asymptotic behaviour of the PDF $P(A , T)$ in the limit of $A, T \rightarrow \infty$ by using the big jump principle. We have extensively discussed and proved in previous section that the continuous stochastic process can be equivalently described by a CTRW, which is a semi-Markovian process, and the statistics of jumps can be analyzed using the tools of renewal theory. The waiting times $\tau$ between jumps events are IID random variables extracted from the PDF \eqref{pdfrenewaltimessolution}, and we found that the average time between jumps $\langle \tau \rangle(v_0)$ is given by the expression \eqref{meanfpt} in the small $v_0$ limit. The individual jumps of our CTRW correspond to the random areas with sign under the excursions $\mathcal{A}$, and in the previous section we found that the tail behaviour of the PDF of these areas is:

\begin{equation}
\label{BJfpassageareas}
f_{v_0}(\mathcal{A}) \underset{\mathcal{A} \rightarrow \infty, v_0 \rightarrow 0}{\sim} v_0 \exp\bigg \{ -\frac{\gamma^{\frac{n+2}{n}}}{\sigma^2} c_n \mathcal{A}^{\frac{2}{n}}  \bigg \}
\end{equation}
The expression of $f_{v_0}({\cal A})$ has been studied in the previous section by fixing ${\cal A}^{1/n}/v_0$ and taking ${\cal A}$ to be large. This means that also $v_0$ is large. However, we expect that in the limit of very large areas, where the ratio ${\cal A}^{1/n}/v_0$ is also large and Equation \eqref{BJfpassageareas} holds, the probability to obtain a certain area ${\cal A}$ is independent of $v_0$. Therefore one should use Equation \eqref{BJfpassageareas} also for small $v_0$. This seems to be confirmed by numerical results in Figure \eqref{fig:ratefunction_full}. 
It is easy to observe that for every $n \great 2$ the exponent of the PDF \eqref{BJfpassageareas} is less than one, so we are dealing with a sub-exponential decay of the PDF of jump lengths. In this context, the single big jump principle finds its natural application \cite{BJ1}. Let us define the PDF that the particle has traveled a total area of value $A$ after a random number $N$ of crossing events $P(A| N)$. The number of crossing events, i.e. our independent and identically distributed jumps, is simply related to the waiting time statistics $\psi(\tau)$ between two crossing events. The big jump principle simply states that the PDF to reach a certain big value of the additive quantity $A$ is simply the PDF to realize a single area under the excursion of size $A$, i.e. a single big jump, times the number of trials in which it can be realized:

\begin{equation}
\label{BJ_A}
P(A|N) \sim N f(A)
\end{equation}

It is easy to conclude that the PDF that the random process arrives in $A$ at time $T$ is given by two main ingredients: the PDF of realizing $N$ crossing events in a total time $T$, i.e. $P(N|T)$ times the PDF of realizing a total area $A$ given $N$ crossing events, i.e. $P(A|N)$. Summing over all possible number of events, one obtain the usual renewal equation, and inserting in it \eqref{BJ_A}, we can describe the tail regime of the PDF, i.e. we return to \eqref{CTRW_BJ}:

\begin{equation}
\label{renewalpdfBJ}
P(A, T) \sim \sum_{N=0}^{\infty} P(A|N)P(N|T) \sim \sum_{N=0}^{\infty}NP(N|T)f(A) \sim \langle N(T) \rangle f(A)
\end{equation}

\begin{figure}
\includegraphics[scale = 0.6]{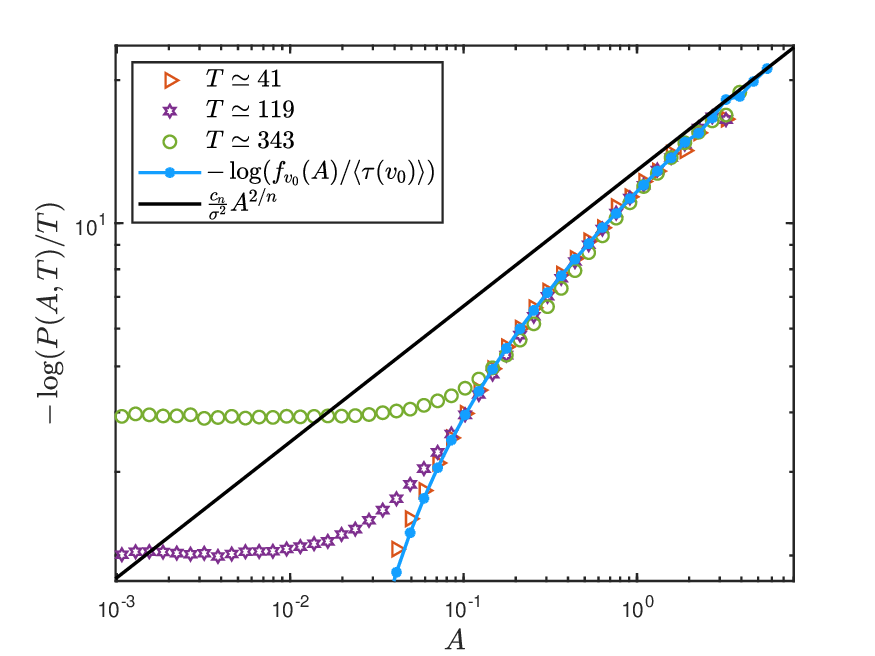}
\caption{Numerical simulations of the tail of the PDF $P(A, T)$ in logarithmic scale performed at different measurement times $T$, represented by the dotted curves, versus its asymptotic estimation through the single big jump principle using the previous obtained results (\ref{CTRW_BJ}, \ref{pdfAgeneral}), represented by the dotted continuous blue curve, in comparison with the analytical value of the far tail limit found in \eqref{coeffcn}, represented by the continuous black line. All simulations were performed for $n=7$, $\gamma=1$ and $\sigma=1$;}
\label{fig:BJ_final}
\end{figure}

\noindent where the PDF of realizing a total area of size $A$ in a time $T$ is simply the PDF of extracting a single big jump of size $A$ from the PDF of single excursions $f(A)$ multiplied by the average number $\langle N(T) \rangle$ of events in the measurement interval in which it can occur. Note that the asymptotic equivalence in \eqref{renewalpdfBJ} is due to the fact that we neglect the contribution of the last excursion $\mathcal{A}^*$ performed in the backward recurrence time $\tau^*$, as discussed in Section \ref{Section2}. We note that this principle is independent of the nature of the PDFs of the waiting times. In fact, the average number of crossing events is simply given by the ratio of the total measurement time to the average time of a single excursion, i.e. $\langle N(T) \rangle = T / \langle \tau \rangle$, and only in the definition of $\langle \tau \rangle$ the distribution of times $\psi(\tau)$ comes into play. This result on the average number of events is trivially true for any PDF of the waiting times from the law of large numbers, as long as we have a finite average time \cite{Burov_N_T}. In our problem this average is strictly related to the mean first passage time calculated in \eqref{meanfpt}. Now, since our key PDFs \eqref{pdfrenewaltimes} and \eqref{BJfpassageareas} are a first passage time and an area under the first passage excursion PDFs starting at an initial velocity $v_0$, we need to take the $v_0 \rightarrow 0$ limit to obtain the renewal process around the zero line, and so we have the big jump result \eqref{CTRW_BJ}:

\begin{equation}
\label{renewalfinalBJ}
P(A, T) \sim \lim_{v_0 \rightarrow 0} \langle N_{v_0}(T) \rangle f_{v_0}(A) \sim \lim_{v_0 \rightarrow 0} \frac{T}{\langle \tau(v_0) \rangle}f_{v_0}(A)
\end{equation}

Substituting expressions \eqref{meanfpt} and \eqref{BJfpassageareas} in the latter equation, we observe that the dependence on the initial condition $v_0$ disappears, since the ratio cancels the initial velocity, and thus we derived the asymptotic behaviour of the full PDF $P(A, T)$ in the long time limit:

\begin{equation}
P(A, T) \underset{A \rightarrow \infty}{\asymp} T \exp\bigg \{ - \frac{\gamma^{\frac{n+2}{n}}}{\sigma^2} c_n A^{\frac{2}{n}}  \bigg \}
\end{equation}


This result coincides with what was found in the instanton approach \cite{TouchetteInst, MeersonInst, NaftaliInst, HarrisInst, AlqahtaniInst, ChenInst}, and the linear dependence of $P(A, T)$ in $T$ is a direct consequence of the law of large numbers, since the possibility of the particle travelling over an area of value $A$ is linearly proportional to the number of trials carried out in a time $T$ to perform a giant jump of that size. As can be observed from the numerical simulations in Figure \ref{fig:BJ_final}, there is very good convergence between the analytical estimate of the far-tail limit and the tails of the $P(A, T)$ for sufficiently large values of $A$, while one can observe the single big jump principle is realised much earlier, as the asymptotic estimate obtained from the big jump principle \eqref{renewalfinalBJ} overlaps with the numerical simulations even at not too large values of $A$. This suggests that the single big jump principle is more general than the analytical result captured by our approach, and this opens the way for interesting discussions and perspectives of future applications.

\section{Conclusions}
\label{Section6}

We have presented an analysis of the anomalous behavior of the rate function that describes the large deviations of the time integrated observable $A=\int_0^T v^n(t) dt$, where $v(t)$ is the Ornstein-Uhlenbeck process \eqref{OUequation}. The rate function is evaluated in two limits: typical fluctuations and ultra rare events. For typical fluctuations, the time-integrated observable behaves as
a Gaussian random variable, analyzed using the Einstein relation \eqref{diffusion} and the Green-Kubo formalism (see Appendix \ref{AppendixB}).
In contrast, rare events are governed by the big jump principle \eqref{CTRW_BJ}, which highlights the dominance of a single extreme excursion rather than the sum of many small steps, as illustrated in Figure \ref{fig:CTRW}.

The big jump principle \cite{BJ1}, applicable to subexponential processes (i.e. for $n>2$), relates the largest extreme excursion to the far tail of the distribution of the time-integrated process.
 Although this principle is established for sums of IID random variables, we extended it to processes with exponentially decaying correlations, such as $v^n(t)$. In this regime, fluctuations exhibit non-diffusive scaling, leading to anomalous behavior in the rate function, consistent with Eq. \eqref{anomalous_rate}. To quantify rare fluctuations using the big jump principle, we exploit an ``excursions technique'' \cite{Barkai2}, including a mapping to the well-known CTRW. The basic approach uses the zero crossings of the Ornstein-Uhlenbeck process, which is a useful strategy since the times between zero crossings are IID random variables. However, the number of zero crossings in an interval of finite duration is either zero or infinity due to the continuous nature of the underlying path. To address this, we study the first-passage time of a regularized process starting at $v_0$ and ending when $v(t)=0$, eventually taking the limit $v_0 \rightarrow 0$. This yields the waiting time distribution in the CTRW language. Furthermore, we calculate the distribution of the random area under the first-passage path, $f_{v_0}(\mathcal{A})$, where the process starts at $v_0$ and ends at $v(t)=0$ for the first time. The big jump principle states that the largest random $\mathcal{A}$ controls the statistics of rare events of the time-integrated random variable $A$. Our challenge was then to find the large deviation theory of $f_{v_0}(\mathcal{A})$.

Thus, the big jump principle highlights the usefulness of studying areas under first-passage functionals \cite{MajumdarBrownianFunctional, KearneyBrownianFunctional, MajumdarFPA}. We find that the rate function of $\mathcal{A}$, Eqs. (\ref{ansatz}, \ref{scalingexpo}), also exhibits anomalous scaling and a critical point. The latter was treated using a non-linear equation for the rate function and tools borrowed from Ref. \cite{BarkaiStretchedExp, BarkaiStretchedExp2}. Interestingly, we identified a connection between the problem of random areas under first-passage functionals for the Ornstein-Uhlenbeck process and a completely different problem \cite{BarkaiStretchedExp, BarkaiStretchedExp2}: the analysis of large deviations of an overdamped particle in a sublinear potential. The latter provides Boltzmann-Gibbs states that are subexponential.
Thus, three rate functions — for time-integrated Ornstein-Uhlenbeck processes, for the area under first-passage functionals, and for overdamped motion in sublinear potentials - are interrelated, a feature partially explained by the big jump principle.

In general, the big jump principle applies to subexponential processes. These include processes with power-tailed distributions, such as the L\'evy walk, as well as processes with stretched exponential statistics, like the case under study. Sampling rare events from data is always a challenge; however, for power-tailed processes, the challenge is relatively easier. For the process under study, sampling issues remain an interesting challenge for future research, requiring techniques that go beyond brute-force sampling \cite{FPTimeOU3,FPTimeOU4, Sampling_conclusions}.

As a final remark, we note that the big jump principle has proven to be an alternative description with respect to the path integral approach, in which the source of the anomalous scaling in the tail regime is identified by an instantonic solution in the weak noise limit \cite{TouchetteInst, MeersonInst, NaftaliInst}. This suggests that there is an important connection with the latter, where big jump effects are the physical sources of these optimal paths. Recent rigorous mathematical works tried to derive with this path integral approach the full rate function of additive variables of other related models, such as Brownian motion with reflections and queueing processes, finding anomalous subexponential scaling in the tail regimes \cite{Instanton_Refl_BM_1, Instanton_Refl_BM_2, Instanton_Refl_BM_3, Instanton_Refl_BM_4, Instanton_Refl_BM_5, Instanton_Refl_BM_6}, and there were also recent analyses of the Ornstein-Uhlenbeck rate function \cite{Math_Instanton_OU_1, Math_Instanton_OU_2}. Our approach may prove to be a new tool to explore the features of these stochastic processes from a different perspective, highlighting the physical source of their fluctuations.

\section{Acknowledgements}

EB thanks the University of Parma for hospitality,
the support of Israel Science Foundation's grant 1614/21 is acknowledged. R. B. is supported by the Project funded under the National Recovery and Resilience Plan (NRRP), Mission 4 Component 2 Investment 1.4—Call for tender No. 3138 of 16/12/2021 of Italian Ministry of University and Research funded by the European Union—NextGenerationEU, Award Number: Project code CN00000023, Concession Decree No. 1033 of 17/06/2022 adopted by the Italian Ministry of University and Research, CUP D93C22000400001,“Sustainable Mobility Center” (CNMS), Spoke 9.

\appendix

\section{The Gaussianity of the Typical Fluctuations in the Renewal Approach}
\label{Appendix0}

In this appendix we prove that the Einstein relation \eqref{Dn} and the Gaussian behaviour of the typical fluctuations of the continuous process \eqref{CLT_P} mapped to a CTRW consisting of the set of renewal random variables $\{ \mathcal{A}_i, \tau_i \}_{i=1}^N$ is valid, even in the case when the variables are not independent, i.e., there are correlations between renewal areas $\mathcal{A}_i$ and renewal times $\tau_i$. The only assumptions necessary to obtain a Gaussian distribution in the bulk of $P(A,T)$ from the renewal approach is the finiteness of $\langle \tau \rangle$ and $\langle \mathcal{A}^2 \rangle$. We will assume for convenience that we are working with symmetric processes, i.e. $n$ odd, whence $\langle \mathcal{A} \rangle = 0$. In general, we define $\phi(A,T)$ as the probability flux that the process reaches an area $A$ at time $T$, and it satisfies the master equation:

\begin{equation}
\phi(A, T) = \delta(A)\delta(T) + \int_{-\infty}^{+\infty} \int_0^T \phi(A-\mathcal{A}, T-\tau)\psi(\mathcal{A},\tau)\ d\mathcal{A} \ d\tau
\end{equation}
where $\psi(\mathcal{A}, \tau)$ is the PDF of extracting a renewal area $\mathcal{A}$ reached a renewal time $\tau$. The PDF $P(A,T)$ that this renewal process reaches a total area $A$ in a time $T$ is also governed by a master equation:

\begin{equation}
P(A,T) = \int_{-\infty}^{\infty} \int_{0}^T \phi(A - \mathcal{A}, T - \mathcal{\tau}) \Psi(\mathcal{A},\tau) \ d\mathcal{A} \ d\tau
\end{equation}
where the survival probability $\Psi(\mathcal{A},\tau)$ can be expressed in term of the conditional probability, i.e. $\Psi(\mathcal{A},\tau)=\Psi(\mathcal{A|\tau})\Psi(\tau)$, and $\Psi(\tau)$ is the survival probability at time $\tau$ introduced in the subsection \ref{subsecfpt}, i.e. $\Psi(\tau) = 1 - \int_0^{\tau} \psi(t) dt = \int_{\tau}^{\infty} \psi(t) dt$, and $\Psi(\mathcal{A}|\tau)$ is the conditional survival PDF of realize an area $\mathcal{A}$ at time $\tau$. Now we switch to Fourier-Laplace space:
\begin{equation}
    \tilde{\phi}(k,s) = \frac{1}{1-\tilde{\psi}(k,s)}
\end{equation}
\begin{equation}
    \label{FLT_Pks}
    \tilde{P}(k,s) = \frac{\tilde{\Psi}(k,s)}{1-\tilde{\psi}(k,s)}
\end{equation}
where $\tilde{\psi}(k,s)$, $\tilde{\phi}(k,s)$, $\tilde{\Psi}(k,s)$ and $\tilde{P}(k,s)$ are respectively the Fourier-Laplace transforms of the PDF of a single renewal $\psi(\mathcal{A}, \tau)$, of the probability flux $\phi(A,T)$, of the survival probability $\Psi(\mathcal{A}, \tau)$ and of the PDF $P(A,T)$. In the small $k,s$ limit one can easily show that the PDF of a single renewal can be written in terms of the first moments of the waiting time $\tau$ and of the renewal area $\mathcal{A}$:
\begin{equation}
    \tilde{\psi}(k,s) = \int_{-\infty}^{+\infty} \int_0^T e^{-s\tau - ik\mathcal{A}} \ \psi(\mathcal{A}, \tau) \ d\mathcal{A} \ d\tau \underset{k,s\rightarrow 0}{\sim}  1 - s\langle \tau \rangle - \frac{k^2}{2}\langle \mathcal{A}^2 \rangle 
\end{equation}
where we can neglect at the first order subleading contributions of higher order moments such as $\langle \tau^2 \rangle$ or $\langle \mathcal{A}\tau \rangle$. We can also show that in the small $k,s$ limit the Fourier-Laplace transform of the survival probability $\tilde{\Psi}(k,s)$ becomes:
\begin{align}
\tilde{\Psi}(k,s) &= \int_{-\infty}^{\infty} \int_{0}^T e^{-s\tau-ik\mathcal{A}} \ \Psi(\mathcal{A},\tau) \ d\mathcal{A} \ d \tau \\
& \underset{k,s\rightarrow 0}{\sim}\int_{-\infty}^{\infty} \Psi(\mathcal{A}|\tau) \ d\mathcal{A} \int_{0}^T  \Psi(\tau) \ d \tau \\
& \underset{k,s\rightarrow 0}{\sim} \langle \tau \rangle
\end{align}
where in the last line integrating over $\mathcal{A}$ the conditional probability gives us its normalization, and the integral of the survival probability over $\tau$ gives us by using integration by parts the first moment of the waiting times. 
Finally, resembling all the results obtained in \eqref{FLT_Pks}, we get that the Fourier-Laplace transform of the original PDF $\tilde{P}(k,s)$ in the small $k,s$ limit is:

\begin{equation}
    \tilde{P}(k,s) \underset{k,s \rightarrow 0}{\sim} \frac{\langle \tau \rangle}{s\langle \tau \rangle + \frac{k^2}{2}\langle \mathcal{A}^2 \rangle}
\end{equation}
which is the Fourier-Laplace transform of a Gaussian distribution. So we finally prove the Gaussianity of the typical fluctuations \eqref{CLT_P} and the Einstein relation \eqref{Dn}.

\section{Derivation of the Backward FPE through Feynman-Kac Formula}
\label{AppendixA}

\noindent In this appendix we derive the equation for the PDF of the area under the first passage excursion $f_{v_0}(\mathcal{A})$ \eqref{pdffirstpassageareas} using a path integral approach based on the Feynman-Kac formula. This derivation is inspired by previous work \cite{KearneyBrownianFunctional, MajumdarBrownianFunctional} for the PDF for the area of the first passage excursion of a generic dynamical observable $U[v(t)]$ in a Brownian motion. Following the same steps, we generalise the derivation to Langevin processes with general force terms. Let us consider the stochastic differential equation (SDE):

\begin{equation}
\label{langeq}
\dot{v}(t)= \mathcal{F}[v(t)] + \sigma \eta(t)    
\end{equation}

\noindent Where $\mathcal{F}[v(t)]$ is a generic force term dependent on the stochastic variable $v(t)$, $\sigma$ is the noise amplitude and $\eta(t)$ is the Gaussian white noise with delta function correlations $\langle \eta(t) \eta(t') \rangle = \delta (t-t')$. The joint probability distribution of a particular path of the variables $\left[ \{ \eta(t) \}, 0 \leq \tau \leq T \right]$ is a Gaussian:

\begin{equation}
\text{Prob}[\{ \eta(t) \}] \sim \text{exp} \left[ -\frac{1}{2} \int_0^T \eta^2(t) dt \right]
\end{equation}

\noindent From the SDE \eqref{langeq} one can express the white noise in terms of $v(t)$ as $\eta(t) = \frac{\dot{v}(t) - \mathcal{F}[v(t)]}{\sigma}$, and thus the joint probability distribution of a path of $[\{ v(t) \}, 0 \leq t \leq T]$ becomes:

\begin{equation}
\text{Prob}[\{ v(t) \}] \sim \text{exp} \left[ -\frac{1}{2\sigma^2} \int_0^T (v(t) - \mathcal{F}[v(t)])^2 dt \right]
\end{equation}

\noindent Now, the joint PDF for a Langevin process to be at velocity $v$ at time $T$ can be written as a general integral transform:

\begin{equation}
P(v,T) = \int_{-\infty}^{+\infty} G(v, T| v_0, 0) P(v_0, 0) dv_0
\end{equation}

\noindent Where $P(v_0, 0)$ is the initial condition and the kernel $G(v, T| v_0, 0)$ is the diffusion propagator, i.e. the probability that any path starting at $v_0$ at time zero evolves to reach velocity $v$ at time $T$, which is naturally represented as a path integral:

\begin{equation}
G(v, t | v_0, 0) =\int_{v(0)=v_0}^{v(T)=v} \mathcal{D} v(t) \exp \left[ -\frac{1}{2\sigma^2} \int_0^T (\dot{v}(t) - \mathcal{F}[v(t)])^2 dt \right]
\end{equation}

\noindent In this formalism, it is easy to insert constraints, in particular  the first-passage property becomes $q(v_0, T)$, i.e. the PDF that any path starting from the initial velocity $v_0 \great 0$ does not collapse to the origin until time $T$. Its path integral representation becomes:

\begin{equation}
q(v_0,T) = \int_{-\infty}^{+\infty} dv \int_{v(0)=v_0}^{v(T)=0} \mathcal{D} v(t) \exp \left[ -\frac{1}{2\sigma^2} \int_0^t (\dot{v}(t) - \mathcal{F}[v(t)])^2 dt \right] \prod_{t=0}^T \theta[v(t)]
\end{equation}

\noindent Here we integrate over all paths starting at $v_0$ and reaching zero velocity in time $T$, and impose through the product of the Heavyside functions $\theta[v(t)]$ that all these paths maintain positive velocity for a time $0 \leq t \leq T$. We can now discuss how to calculate the statistical properties of the first passage functional $\mathcal{A}$, defined as:

\begin{equation}
\mathcal{A} = \int_{0}^{\tau} U[v(t)] dt
\end{equation}

\noindent where $v(t)$ is the generic Langevin path governed by the initial SDE \eqref{langeq} that starts at $v_0 \great 0$ at time zero and propagates till the first passage time $\tau$, where it first crosses the origin. As long as $\mathcal{A}$ has positive support, if we want to compute the PDF $f_{v_0}(\mathcal{A})$, i.e. the PDF that any path of $v(t)$ reaches the value of $\mathcal{A}$ starting from a certain initial condition $v_0 \great 0$ before crossing for the first time the origin, it is useful to consider its Laplace transform $Q_{v_0}(s)$, where $s$ is the dual variable associated with $\mathcal{A}$. At this point the Feynman-Kac formula comes to our help, and it states that the Laplace transform of $f_{v_0}(\mathcal{A})$ is the expectation value over all possible first-passage Langevin paths starting from the initial velocity $v_0$. Hence, we have that:

\begin{align}
\label{feynmankacfp}
Q_{v_0}(s) &= \int_0^{+\infty} e^{-s \mathcal{A}} f_{v_0}(\mathcal{A}) d\mathcal{A} = \langle e^{-s\int_0^{\tau} U[v(t)]dt} \rangle_{v_0}\\
&= \int_{0}^{+\infty} dv \int_{0}^{+\infty} d\tau \int_{v(0)=v_0}^{v(\tau)=0} \mathcal{D} v(t) \exp \left[ - \int_0^{\tau} \left( \frac{1}{2\sigma^2}(\dot{v}(t) - \mathcal{F}[v(t)])^2 +s U[v(t)] \right) dt \right] \prod_{\tau=0}^{\tau} \theta[v(t)]
\end{align}

\noindent Now, the trick we use to build an evolution equation for the Laplace transformed PDF $Q_{v_0}(s)$ is to divide the integration interval into two subintervals $[0, \tau] = [0, \Delta t] \cup [\Delta t, \tau]$, with $\Delta t$ small. Returning to the initial SDE \eqref{langeq} we know that in a small time interval $\Delta t$ the velocity increment is equal to $\Delta v = \mathcal{F}[v(0)]\Delta t + \sigma \eta(0) \Delta t$,  and note that $v(0)=v_0$. Also the integral for $\mathcal{A}$ is splitted in two parts, and since the initial value is $v_0$, one can assume for $\Delta t$ small enough that $\int_{0}^{\Delta t} U[v(t)] dt = U(v_0)\Delta t$. Then equation \eqref{feynmankacfp} can be written as:

\begin{equation}
Q_{v_0}(s) = \langle e^{-sU(v_0)\Delta t} e^{-s\int_{\Delta t}^{\tau} U[v(t)] dt}   \rangle_{v_0} = \langle e^{-sU(v_0)\Delta t} Q_{v_0+\Delta v}(s)\rangle_{\Delta v}
\end{equation}

\noindent Expanding the first exponential by $\Delta t$ small at first order we have that $e^{-s U(v_0) \Delta t} \sim 1 - s U(v_0) \Delta t$, and assuming also $\Delta v$ small, the second expression can be expanded in Taylor series, so $Q_{v_0+\Delta v}(s) \sim Q_{v_0}(s) + \frac{\partial Q}{\partial v_0}\Delta v +\frac{1}{2} \frac{\partial^2 Q}{\partial v_0^2}\Delta v^2$. Finally, resembling all together and averaging over $\Delta t$, we have that:

\begin{align}
Q_{v_0}(s) =& \ \bigg \langle (1-sU(v_0)\Delta t) \left( Q_{v_0}(s) + \frac{\partial Q}{\partial v_0}\Delta v +\frac{1}{2} \frac{\partial^2 Q}{\partial v_0^2}\Delta v^2 \right) \bigg \rangle_{\Delta t} \\
=& \ \bigg \langle (1-sU(v_0)\Delta t) \left( Q_{v_0}(s) + \frac{\partial Q}{\partial v_0}(\mathcal{F}[v(0)]\Delta t + \sigma \eta(0) \Delta t) +\frac{1}{2} \frac{\partial^2 Q}{\partial v_0^2}(\mathcal{F}[v(0)]\Delta t + \sigma \eta(0) \Delta t)^2  \right) \bigg \rangle_{\Delta t} \\
=& \ Q_{v_0}(s) -sU(v_0)Q_{v_0}(s)\Delta t + \mathcal{F}(v_0)\frac{\partial Q}{\partial v_0}\Delta t + \frac{\partial Q}{\partial v_0} \sigma \langle \eta(0) \rangle_{\Delta t} \Delta t + \frac{1}{2}\mathcal{F}(v_0)^2 \frac{\partial^2 Q}{\partial v_0^2} \Delta t^2  \\
& + \mathcal{F}(v_0) \frac{\partial^2 Q}{\partial v_0} \sigma \langle \eta(0) \rangle_{\Delta t} \Delta t^2 + \frac{1}{2}\frac{\partial^2 Q}{\partial v_0^2} \sigma^2 \langle \eta^2(0) \rangle_{\Delta t} \Delta t^2 + o(\Delta t^2) \nonumber
\end{align}

\noindent From the properties of white noise we know that $\langle \eta(0) \rangle_{\Delta t} = 0$ and in the limit of $\Delta t$ small $\langle \eta(0)^2 \rangle_{\Delta t} =\frac{1}{\Delta t}$, so we have that, simplifying the common terms and neglecting the terms of order $\Delta t^2$ and higher, we obtain:

\begin{equation}
\left( \frac{\sigma^2}{2} \frac{\partial^2 Q_{v_0}(s)}{\partial v_0^2} + \mathcal{F}(v_0) \frac{\partial Q_{v_0}(s)}{\partial v_0} - s U(v_0) Q_{v_0}(s)\right) \Delta t + o(\Delta t^2) = 0 
\end{equation}

\noindent Thus, the differential equation for the Laplace functional $Q_{v_0}(s)$ at the leading order in $\Delta t$ is:

\begin{equation}
 \frac{\sigma^2}{2} \frac{\partial^2 Q(s,v_0)}{\partial v_0^2} + \mathcal{F}(v_0) \frac{\partial Q_{v_0}(s)}{\partial v_0} - s U(v_0) Q_{v_0}(s)=0
\end{equation}

\noindent Finally we return to real space by performing the inverse Laplace transform to obtain an equation for $f_{v_0}(\mathcal{A})$, where we use the initial condition $f_{v_0}(\mathcal{A}=0)=0$ since it is not possible to cover a null area as long as $v_0 \great 0$. So we finally derived the general backward equation for the first passage functional $f_{v_0}(\mathcal{A})$: 

\begin{equation}
 \frac{\sigma^2}{2} \frac{\partial^2 f_{v_0}(\mathcal{A})}{\partial v_0^2} + \mathcal{F}(v_0) \frac{\partial f_{v_0}(\mathcal{A})}{\partial v_0} - U(v_0) \frac{\partial f_{v_0}(\mathcal{A})}{\partial \mathcal{A}}=0
\end{equation}

\noindent In our case study we consider the Ornstein-Uhlenbeck process governed by the equation \eqref{OUequation} from which we easily recognise that $\mathcal{F}(v_0) = -\gamma v_0$, and in particular the first passage functional of our interest is $\mathcal{A}=\int_0^{\tau} v^n(t) dt$, whence $U(v_0)=v_0^n$. Then, by finally substituting the appropriate terms, we return to our desired backward equation \eqref{pdffirstpassageareas};

\section{Velocity Correlation Function and Green-Kubo Formula}
\label{AppendixB}

\noindent We start our calculation from the already known results on the one-dimensional Ornstein-Uhlenbeck process for $v(t)$. The complete expression of its joint PDF $P(v, t | v_0, 0)$, i.e. the PDF that a single path starting from $v_0$ at time zero reaches a velocity $v$ at time $t$ is:

\begin{equation}
\label{timedeppdf}
P(v, t | v_0, 0) = \frac{1}{\sqrt{2 \pi \text{Var}(v)}} \exp \left( -\frac{(v-\langle v \rangle)^2}{2 \text{Var}(v)} \right)
\end{equation}

\noindent Where $\langle v \rangle = v_0 e^{-\gamma t}$ and  $\text{Var}(v) = \frac{\sigma^2}{2\gamma}(1-e^{-2\gamma t})$. Now, in the long time limit $t \rightarrow \infty$ the joint PDF of the Ornstein-Uhlenbeck process $P(v, t | v_0, 0)$ goes to a stationary solution $P_{eq}(v)$, which is again a Gaussian, but its mean and variance are now independent from the initial condition, so $\langle v \rangle_{eq}=0$ and $\text{Var}(v)_{eq}=\frac{\sigma^2}{2\gamma}$. In the calculation of the two-point correlation function we have to distinguish two types of averages: let us denote with $\langle ... \rangle$ the mean values computed with respect to time dependent PDF \eqref{timedeppdf} and with $\langle ... \rangle_{eq}$ the mean values computed with respect to the time-independent equilibrium PDF $P_{eq}(v)$. We are interested in computing the $n$-th moment of $v^n$ in both cases. Let us start with the time-dependent case:

\begin{equation}
\langle v^n(t) \rangle = \int_{-\infty}^{+\infty} v^n  P(v, t | v_0, 0) dv = \int_{-\infty}^{+\infty} \frac{v^n}{\sqrt{2\pi \text{Var}(v)}} e^{-\frac{(v-\langle v \rangle)^2}{2 \text{Var}(v)}} dv 
\end{equation}

\noindent To calculate this integral we use a simple change of variable $\mathcal{V} = v - \langle v \rangle$, so we have that $v^n = (\langle v \rangle + \mathcal{V})^n$, and this binomial of degree $n$ can be developed by Newton's formula $(a+b)^n = \sum_{k=0}^n \binom{n}{k} a^{k}b^{n-k}$:

\begin{align}
\langle v^n(t) \rangle &= \int_{-\infty}^{+\infty} (\langle v \rangle + \mathcal{V})^n \frac{1}{\sqrt{2\pi \text{Var}(v)}} e^{-\frac{\mathcal{V}^2}{2 \text{Var}(v)}} d\mathcal{V} \\ 
&= \int_{-\infty}^{+\infty} \sum_{k=0}^n \binom{n}{k} \langle v \rangle^k \mathcal{V}^{n-k} \frac{1}{\sqrt{2\pi \text{Var}(v)}} e^{-\frac{\mathcal{V}^2}{2 \text{Var}(v)}} d\mathcal{V} \\  
&= \sum_{k=0}^n \binom{n}{k} \langle v \rangle^k \int_{-\infty}^{+\infty} \frac{\mathcal{V}^{n-k}}{\sqrt{2\pi \text{Var}(v)}} e^{-\frac{\mathcal{V}^2}{2 \text{Var}(v)}} d\mathcal{V} \\ 
&= \sum_{k=0}^n \binom{n}{k} \langle v \rangle^k \frac{2^{\frac{n-k}{2}} \Gamma \left( \frac{n-k+1}{2} \right)}{\sqrt{\pi}} \text{Var}(v)^{\frac{n-k}{2}} \\ 
&= \sum_{k=0}^n \binom{n}{k} \alpha_{n-k} v_0^k \left(\frac{\sigma^2}{2\gamma} \right)^{\frac{n-k}{2}} \left( 1-e^{-2\gamma t}\right)^{\frac{n-k}{2}} e^{-k\gamma t}
\end{align}

Now let's calculate the mean value of $v^n$ at equilibrium. In this case, the result is a simple Gaussian integral:

\begin{align}
\langle v^n \rangle_{eq} = \int_{-\infty}^{+\infty} v^n P_{eq}(v) dv &= \int_{-\infty}^{+\infty} \frac{v^n}{\sqrt{2 \pi \text{Var}(v)_{eq}}} e^{-\frac{(v-\langle v \rangle_{eq})^2}{2\text{Var}(v)_{eq}}} dv \\ 
&=\int_{-\infty}^{+\infty} \sqrt{\frac{\pi}{\gamma}}\frac{1}{\sigma} v^n e^{-\frac{\gamma v2}{\sigma^2}} dv \\ 
&= \left( \frac{\sigma^2}{2\gamma} \right)^{\frac{n}{2}} \alpha_n
\end{align}

\noindent Thus, we can now calculate the two-point correlation function of the velocity. In particular, in order to include the odd and even cases in a single expression, we study the correlation function for the shifted variable $\Delta v^n = v^n - \langle v_0^n \rangle_{eq}$, where $v_0=v(0)$. We know that from general properties of stationary processes $\text{corr}_n(t_1, t_2) = \langle \Delta v^n(t_1) \Delta v^n(t_2) \rangle = \langle \Delta v^n(t_2 + t) \Delta v^n (t_2) \rangle$, and so redefining $t=|t_1-t_2|$, and assuming that time $t$ is sufficiently large to bring the process to equilibrium, we have:

\begin{align}
\text{corr}_n(t) &= \langle \Delta v^n(t) \Delta v^n(0) \rangle_{eq} = \langle \langle v^n(t) \rangle v_0^n \rangle_{eq} - \langle \langle v^n(t) \rangle \langle v_0^n \rangle_{eq} \rangle_{eq} \\
&= \sum_{k=0}^n \binom{n}{k} \alpha_{n-k} \left( \frac{\sigma^2}{2\gamma} \right)^{\frac{n-k}{2}} (1 - e^{-2\gamma t})^{\frac{n-k}{2}} e^{-k \gamma t} \left[ \langle v_0^{n+k} \rangle_{eq} - \langle v_0^n\rangle_{eq} \langle v_0^k \rangle_{eq} \right] \\ 
&=  \frac{\sigma^{2n}}{2^n \gamma^{n+1}}  \sum_{k=0}^n \binom{n}{k} \alpha_{n-k}(\alpha_{n+k}-\alpha_{n}\alpha_k) e^{-k\gamma t} (1 - e^{-2\gamma t})^{\frac{n-k}{2}}
\end{align}

\noindent Finally we can apply Green-Kubo's formula integrating the correlation function over time, and to integrate the binomial of degree $\frac{n-k}{2}$ we again use Newton's formula:

\begin{align}
D_n &= \int_{0}^{\infty} \text{corr}_n(t) dt \\ 
&= \frac{\sigma^{2n}}{2^n \gamma^{n+1}}  \sum_{k=0}^n \binom{n}{k} \alpha_{n-k}(\alpha_{n+k}-\alpha_{n}\alpha_k) \int_0^{+\infty} e^{-k\gamma t} (1 - e^{-2\gamma t})^{\frac{n-k}{2}} dt \\ 
&= \frac{\sigma^{2n}}{2^n \gamma^{n+1}}  \sum_{k=0}^n \binom{n}{k} \alpha_{n-k}(\alpha_{n+k}-\alpha_{n}\alpha_k) \int_0^{+\infty} e^{-k\gamma t} \sum_{m=0}^{\frac{n-k}{2}} \binom{\frac{n-k}{2}}{m} (-1)^{\frac{n-k}{2}-m} e^{-(n-k-2m)\gamma t} dt \\ 
&=\frac{\sigma^{2n}}{2^n \gamma^{n+1}}  \sum_{k=0}^n \binom{n}{k} \alpha_{n-k}(\alpha_{n+k}-\alpha_{n}\alpha_k) \sum_{m=0}^{\frac{n-k}{2}} \binom{\frac{n-k}{2}}{m} (-1)^{\frac{n-k}{2}-m} \int_0^{+\infty} e^{-(n-2m)\gamma t} dt \\
&= \frac{\sigma^{2n}}{2^n \gamma^{n+1}}  \sum_{k=0}^n \binom{n}{k} \alpha_{n-k}(\alpha_{n+k}-\alpha_{n}\alpha_k) \left(\sum_{m=0}^{\frac{n-k}{2}} \binom{\frac{n-k}{2}}{m}  \frac{(-1)^{\frac{n-k}{2}-m}}{n-2m} \right) 
\end{align}

\noindent So we derived the an alternative expression for the diffusion constant $D_n$ \eqref{Dn} and it can be checked in a simple way that this expression is valid both in the case of $n$ even and $n$ odd and it's equal to the result obtained with the CTRW approach. If $n$ is odd we have that $\alpha_n=0$, so the second term in the last line disappears.

\section{Mean First Passage Time and Mean Squared First Passage Excursion}
\label{AppendixC}

\subsection{Mean First Passage Time}

\noindent We start from the equation \eqref{equationmfpt} and recall the absorbing boundary condition at the line $\langle \tau \rangle(v_0 \rightarrow 0) = 0$ and the reflective boundary condition at infinity $\partial_{v_0} \langle \tau \rangle(v_0 \rightarrow \infty) = 0$. To solve this second order ordinary differential equation, one can reduce it to an equation of first order by defining $y(v_0) = \partial_{v_0} \langle \tau \rangle (v_0)$, and this is obtained:

\begin{equation}
\frac{\sigma^2}{2} \frac{\partial}{\partial v_0} y(v_0) - \gamma v_0 y(v_0) = -1 
\end{equation}

\noindent where the reflective boundary condition for this equation becomes $y(v_0 \rightarrow \infty) = 0$, from which it follows that:

\begin{align}
y(v_0) &= -\frac{2}{\sigma} e^{\frac{\gamma v_0^2}{\sigma^2}} \int_{v_0}^{\infty} e^{-\frac{\gamma v^2}{\sigma^2}} dv\\ \nonumber
&= \frac{1}{\sigma} \sqrt{\frac{\pi}{\gamma}} e^{\frac{\gamma v_0^2}{\sigma^2}} \left( 1 -\text{erf}\left( \frac{\sqrt{\gamma} v_0}{\sigma} \right) \right)
\end{align}

\noindent At this point, to obtain the mean first passage time $\langle \tau \rangle (v_0)$ it is sufficient to integrate $y(v_0)$ and impose the absorbing boundary condition $y(v_0 \rightarrow 0) = 0$ at the extremes of integration. So we get:

\begin{align}
\langle \tau \rangle (v_0) &= \int_0^{v_0} y(v) dv = \int_0^{v_0} \frac{\partial \langle \tau \rangle (v)}{\partial v} dv \\ \nonumber
&= \frac{1}{\sigma} \sqrt{\frac{\pi}{\gamma}} \int_0^{v_0} e^{\frac{\gamma v^2}{\sigma^2}} \left( 1 -\text{erf}\left( \frac{\sqrt{\gamma} v}{\sigma} \right) \right) dv \\ \nonumber 
&= \frac{\pi}{2 \gamma} \text{erfi}\left( \frac{\sqrt{\gamma}v_0}{\sigma} \right) - \frac{1}{\sigma^2} v_0^2 \ {}_2F_2 \left( 1, 1; \frac{3}{2}, 2; \frac{\gamma v_0^2}{\sigma^2} \right) 
\end{align}

\noindent Then, having obtained the complete expression of $\langle \tau \rangle (v_0)$ for any initial condition $v_0$, we calculate its asymptotic expression in the limit of $v_0 \rightarrow 0$. It is known that ${}_2F_2\left( 1,1 ; \frac{3}{2}, 2 ; \frac{\gamma v_0^2}{\sigma^2} \right) \sim 1 + o(v_0^2)$ and $\text{erfi}\left(\frac{\sqrt{\gamma}v_0}{\sigma}\right) \sim 2\sqrt{\frac{\gamma}{\pi}} \frac{1}{\sigma} v_0 + o(v_0^2)$. Then, putting everything back together, the leading order of $\langle \tau \rangle (v_0)$ for $v_0 \rightarrow 0$ turns out to be:

\begin{align}
\langle \tau \rangle (v_0)  &\underset{v_0 \rightarrow 0}{\sim} \frac{\pi}{2 \gamma} \left(2\sqrt{\frac{\gamma}{\pi}} \frac{1}{\sigma} v_0\right) - \frac{1}{\sigma^2} v_0^2 \\ \nonumber
&\underset{v_0 \rightarrow 0}{\sim}\frac{1}{\sigma} \sqrt{\frac{\pi}{\gamma}} v_0
\end{align}

\subsection{Mean Squared First Passage Excursion for $n$ odd}

\noindent Starting again from the equation for the mean first passage excursion \eqref{equationmfpexcursion} and recalling the absorbing boundary condition at the lower bound $\langle \mathcal{A} \rangle (v_0 \rightarrow 0) = 0$ and the reflective boundary condition at infinity $\partial_{v_0}, \langle \mathcal{A} \rangle (v_0 \rightarrow \infty) = 0$ we now evaluate it in the case of $n$ odd. Once again, we can define the first derivative $y(v_0) = \partial_{v_0} \langle \mathcal{A} \rangle$ and solve the associated first order ODE by variation of constants method. The equation becomes:

\begin{equation}
\frac{\sigma^2}{2} \frac{\partial}{\partial v_0} y(v_0) - \gamma v_0 y(v_0) - 2v_0^n = 0
\end{equation}

\noindent Imposing the reflective boundary condition $y(v_0 \rightarrow \infty) = 0$ and performing the change of variable to $x=\frac{\sqrt{2\gamma} v_0}{\sigma}$ we obtain the following Gaussian integral \cite{Korotkov}:

\begin{align}
y(v_0) &= \frac{4}{\sigma^2} e^{\frac{\gamma v_0^2}{\sigma^2}} \int_{v_0}^{\infty} v^n e^{-\frac{\gamma v^2}{\sigma^2}} dv \\ \nonumber
&= 2 \frac{\sigma^{n-1}}{2^{\frac{n-1}{2}}\gamma^{\frac{n+1}{2}}} e^{\frac{\gamma v_0^2}{\sigma^2}} \int_{\frac{\sqrt{2\gamma} v_0}{\sigma}}^{\infty} x^n e^{-\frac{1}{2}x^2} dx \\ \nonumber
&= 2 \frac{\sigma^{n-1}}{2^{\frac{n-1}{2}}\gamma^{\frac{n+1}{2}}} \sum_{j=0}^{\frac{n-1}{2}} \frac{(n-1)!!}{(2j)!!} \frac{2^j \gamma^j}{\sigma^{2j}} v_0^{2j} 
\end{align}

\noindent Now, integrating over $v_0$ and imposing the absorbing boundary condition $\langle \mathcal{A} \rangle(v_0 = 0) = 0$ we get:

\begin{align}
\langle \mathcal{A} \rangle (v_0) &= \int_0^{v_0} y(v) dv = \int_0^{v_0} \frac{\partial \langle \mathcal{A} \rangle (v)}{\partial v} dv \\ \nonumber
&= 2 \frac{\sigma^{n-1}}{2^{\frac{n-1}{2}}\gamma^{\frac{n+1}{2}}} \sum_{j=0}^{\frac{n-1}{2}} \frac{(n-1)!!}{(2j)!!} \frac{2^j \gamma^j}{\sigma^{2j}} \int_0^{v_0} v^{2j} dv \\ \nonumber
&= 2 \frac{\sigma^{n-1}}{2^{\frac{n-1}{2}}\gamma^{\frac{n+1}{2}}} \sum_{j=0}^{\frac{n-1}{2}} \frac{(n-1)!!}{(2j)!!} \frac{2^j \gamma^j}{\sigma^{2j}} \frac{v_0^{2j+1}}{2j+1}
\end{align}

\noindent It is now possible to recursively solve the backward equation for the mean squared excursion $\langle \mathcal{A}^2 \rangle(v_0)$ \eqref{equationmeansquaredexc} from the solution just found using the asymptotic expression of $\langle \mathcal{A}^2 \rangle(v_0)$ for $v_0 \rightarrow 0$, and then we finally derive that:

\begin{align}
\langle \mathcal{A}^2 \rangle (v_0)  &\underset{v_0 \rightarrow 0}{\sim} 2 v_0 \int_{0}^{+\infty} v^n e^{-\frac{\gamma v^2}{\sigma^2}} \langle \mathcal{A} \rangle(v) dv \\ \nonumber
&\underset{v_0 \rightarrow 0}{\sim} 4 v_0 \frac{\sigma^{n-1}}{2^{\frac{n-1}{2}}\gamma^{\frac{n+1}{2}}} \sum_{j=0}^{\frac{n-1}{2}} \frac{(n-1)!!}{(2j)!!} \frac{2^j \gamma^j}{\sigma^{2j}} \frac{1}{2j+1} \int_0^{\infty} v_0^{n+2j+1} e^{-\frac{\gamma v_0^2}{\sigma^2}} dv_0
\end{align}

\noindent At this point, to solve the Gaussian integral we use the well-known identity $\int_0^{\infty} x^n e^{-ax^2} = \frac{(n-1)!!}{2^{\frac{n}{2}+1} a^{\frac{n}{2}}} \sqrt{\frac{\pi}{a}}$ and remember that for small $v_0$ the mean first passage time is $\langle \tau \rangle (v_0) \sim \sqrt{\frac{\pi}{\gamma}} \sigma v_0$ we obtain that:

\begin{align}
\langle \mathcal{A}^2 \rangle_n (v_0)  &\underset{v_0 \rightarrow 0}{\sim} 2 \frac{\sigma^{2n}}{2^n\gamma^{n+1}} \sum_{j=0}^{\frac{n-1}{2}} \frac{(n-1)!!}{(2j)!!}\frac{(n+2j)!!}{2j+1} \sigma \sqrt{\frac{\pi}{\gamma}} v_0 \\ \nonumber
&\underset{v_0 \rightarrow 0}{\sim} 2 \frac{\sigma^{2n}}{2^n\gamma^{n+1}} \sum_{j=0}^{\frac{n-1}{2}} \frac{(n-1)!!}{(2j)!!}\frac{(n+2j)!!}{2j+1} \langle \tau \rangle (v_0)
\end{align}

\subsection{Mean Squared First Passage Excursion for $n$ even}

\noindent The calculation of the mean squared displacement $\langle \mathcal{A}^2 \rangle$ in the case of even $n$ is very similar to the case of odd $n$, with one important difference to consider: in this process the mean value of the dynamical observable $\langle \mathcal{A} \rangle(v_0) \neq 0$ as $v_0 \rightarrow 0$, and if the initial condition of the system is for example given by $v_0 \great 0$, the integral of the even function will always be positive since the integrand function will always be positive for parity, and consequently it will never be possible to observe the process return towards the zero line. In other words, the Gaussian process  has the mean value shifted to a non-zero value $\langle \mathcal{A} \rangle$, and therefore, if we want to study the first passage excursion with respect to the zero line, it is necessary to study the shifted observable $\mathcal{A}' = \mathcal{A} - \langle \mathcal{A} \rangle$. To calculate this expectation value, we can remember the expression of the mean value at equilibrium of $\langle v^n \rangle_{eq} = \alpha_n$, with $\alpha_n = \left(\frac{\sigma^2}{2\gamma}\right)^{n/2} \frac{2^{n/2} \Gamma \left( \frac{n+1}{2} \right)}{\sqrt{\pi}}$, then the backward equation for the mean value $\langle \mathcal{A'} \rangle$ becomes:

\begin{equation}
-\gamma v_0 \frac{\partial}{\partial v_0} \langle \mathcal{A'} \rangle (v_0) + \frac{\sigma^2}{2} \frac{\partial^2}{\partial v_0^2} \langle \mathcal{A'} \rangle (v_0) = 2(v_0^n - \alpha_n)
\end{equation}

\noindent The boundary conditions for this equation remain the same as for the odd $n$ case, so it can be solved in the same way as before, by first solving the equation for $y(v_0) = \partial_{v_0} \langle \mathcal{A'} \rangle$, imposing the reflective boundary condition $y(v_0 \rightarrow \infty)=0$ and using the method of variation of constants:

\begin{align}
y(v_0) &= \frac{4}{\sigma^2} e^{\frac{\gamma v_0^2}{\sigma^2}} \int_{v_0}^{\infty} (v^n - \alpha_n) e^{-\frac{\gamma v^2}{\sigma^2}} dv \\ \nonumber
&= 2 \frac{\sigma^{n-1}}{2^{\frac{n-1}{2}}\gamma^{\frac{n+1}{2}}} e^{\frac{\gamma v_0^2}{\sigma^2}} \left( \int_{\frac{\sqrt{2\gamma} v_0}{\sigma}}^{\infty} x^n e^{-\frac{1}{2}x^2} dx - \alpha_n \int_{\frac{\sqrt{2\gamma} v_0}{\sigma}}^{\infty} e^{-\frac{1}{2}x^2} dx \right) \\ \nonumber
&= 2 \frac{\sigma^{n-1}}{2^{\frac{n-1}{2}}\gamma^{\frac{n+1}{2}}} \sum_{j=0}^{\frac{n}{2}-1} \frac{(n-1)!!}{(2j+1)!!} \frac{2^{j+\frac{1}{2}} \gamma^{j+\frac{1}{2}}}{\sigma^{2j+1}} v_0^{2j+1} 
\end{align}

\noindent Again, we integrate over $v_0$ and fix the absorbing boundary condition $\langle \mathcal{A'} \rangle(v_0 = 0) = 0$:

\begin{align}
\langle \mathcal{A'} \rangle_n (v_0) &= \int_0^{v_0} y(v) dv = \int_0^{v_0} \frac{\partial \langle \mathcal{A'} \rangle_n (v)}{\partial v} dv \\ \nonumber
&= 2 \frac{\sigma^{n-1}}{2^{\frac{n-1}{2}}\gamma^{\frac{n+1}{2}}} \sum_{j=0}^{\frac{n}{2}-1} \frac{(n-1)!!}{(2j+1)!!} \frac{2^{j+\frac{1}{2}} \gamma^{j+\frac{1}{2}}}{\sigma^{2j+1}} \frac{v_0^{2j+2}}{2j+2}
\end{align}

\begin{table}
\begin{tabular}{@{}|c|c|c|@{}}
\toprule
\hline
  & \textbf{$n$ odd} & \textbf{$n$ even} \\ \midrule
\hline
\textbf{RT} & $\sum_{j=0}^{\frac{n-1}{2}} \frac{(n-1)!!}{2j!!} \frac{(n+2j)!!}{2j+1}$ & $ \sum_{j=0}^{\frac{n}{2}-1} \frac{(n-1)!!}{(2j+1)!!} \frac{\left((n+2j+1)!! - (n-1)!!(2j+1)!! \right)}{2j+2}$   \\
\hline
\textbf{GK} & $ \sum_{k=0}^n \binom{n}{k} \alpha_{n-k} \alpha_{n+k} \left( \sum_{m=0}^{\frac{n-k}{2}} \binom{\frac{n-k}{2}}{m} \frac{(-1)^{\frac{n-k}{2}-m}}{n-2m} \right)$  & $\sum_{k=0}^n \binom{n}{k} \alpha_{n-k} ( \alpha_{n+k} - \alpha_n \alpha_k) \left( \sum_{m=0}^{\frac{n-k}{2}} \binom{\frac{n-k}{2}}{m} \frac{(-1)^{\frac{n-k}{2}-m}}{n-2m} \right)$ \\
\hline
\rule[-5mm]{0mm}{0.7cm}
\textbf{DV}\cite{NaftaliInst} & \multicolumn{2}{|c|}{\multirow{2}{*}{$\left( 4 \sum_{m=1}^{+\infty} \frac{|\langle m | v^n | 0 \rangle|^2}{E_m - E_0} \right)^{-1}$}} \\
\hline
\bottomrule
\end{tabular}
\caption{Summary of analytical results of the diffusion coefficient $D_n$ normalised with respect to the dimensional prefactor $\frac{\sigma^{2n}}{2^n \gamma^{n+1}}$ calculated using different approaches. RT stands for renewal theory approach used in Section \ref{Section3}, GK stands for Green-Kubo approach computed in Appendix \ref{AppendixB}, and DV stands for D$\ddot{o}$nsker-Varadhan approach, used in \cite{NaftaliInst}. All analytical expressions found by these different methods are formally equivalent, and have been tested numerically for several values of $n$;}
\label{tab:caption}
\end{table}

\noindent At this point it is also easy to see how the equation for the mean squared excursion $\langle \mathcal{A'}^2 \rangle(v_0)$ will also be modified:

\begin{equation}
-\gamma v_0 \frac{\partial}{\partial v_0} \langle \mathcal{A'}^2 \rangle(v_0) + \frac{\sigma^2}{2} \frac{\partial^2}{\partial v_0^2} \langle \mathcal{A'}^2 \rangle (v_0) = 2(v_0^n - \alpha_n)\langle \mathcal{A'} \rangle (v_0)
\end{equation}

\noindent from which we derive in the same way for $n$ odd the asymptotic expression for the mean squared excursion in the limit of $v_0 \rightarrow 0$, which is the result we are interested in:

\begin{align}
\langle \mathcal{A'}^2 \rangle (v_0)  &\underset{v_0 \rightarrow 0}{\sim} 2 v_0 \int_{0}^{+\infty} (v^n - \alpha_n) e^{-\frac{\gamma v^2}{\sigma^2}} \langle \mathcal{A} \rangle(v) dv \\ \nonumber
&\underset{v_0 \rightarrow 0}{\sim} 4 v_0 \frac{\sigma^{n-1}}{2^{\frac{n-1}{2}}\gamma^{\frac{n+1}{2}}} \sum_{j=0}^{\frac{n}{2}-1} \frac{(n-1)!!}{(2j+1)!!} \frac{2^{j+\frac{1}{2}} \gamma^{j+\frac{1}{2}}}{\sigma^{2j+1}} \frac{1}{2j+2} \left(\int_0^{\infty} v_0^{n+2j+2} e^{-\frac{\gamma v_0^2}{\sigma^2}} dv_0 - \alpha_n \int_0^{\infty} v_0^{2j+2} e^{-\frac{\gamma v_0^2}{\sigma^2}} dv_0 \right) \\ \nonumber
&\underset{v_0 \rightarrow 0}{\sim} 2 \frac{\sigma^{2n}}{2^n\gamma^{n+1}} \sum_{j=0}^{\frac{n}{2}-1} \frac{(n-1)!!}{(2j+1)!!}\frac{(n+2j+1)!!-(n-1)!!(2j+1)!!}{2j+2} \sigma \sqrt{\frac{\pi}{\gamma}} v_0 \\ \nonumber
&\underset{v_0 \rightarrow 0}{\sim} 2 \frac{\sigma^{2n}}{2^n\gamma^{n+1}} \sum_{j=0}^{\frac{n}{2}-1} \frac{(n-1)!!}{(2j+1)!!}\frac{(n+2j+1)!!-(n-1)!!(2j+1)!!}{2j+2} \langle \tau \rangle (v_0)
\end{align}

The various results found for $D_n$ by different methods are summarised in Table \ref{tab:caption};

\section{Calculation of the Rate Function Equation}
\label{AppendixD}

\noindent In this appendix we show how the backward equation for the PDF $f_{v_0}(\mathcal{A})$ \eqref{pdffirstpassageareas} can be mapped into an ODE for an exponential rate function \eqref{ratefunctionequation}. Our starting point is the exponential ansatz assigned in \eqref{ansatz}, and we now see how the various terms in the equation \eqref{pdffirstpassageareas} are transformed:

\begin{itemize}
    \item 
    $\begin{aligned}
    v_0^n \frac{\partial f_{v_0}(\mathcal{A})}{\partial \mathcal{A}} &= v_0^n \frac{\partial}{\partial \mathcal{A}} \left(  v_0 \mathcal{A}^{\xi} \exp \bigg \{ - \frac{\mathcal{A}^{\delta}}{\sigma^2}  \mathcal{I}\left(\frac{v_0}{\mathcal{A}^{\beta}}\right) \bigg \}  \right) \\
    &= v_0^{n+1} \mathcal{A}^{\xi}\exp \bigg \{ - \frac{\mathcal{A}^{\delta}}{\sigma^2}  \mathcal{I}\left(\frac{v_0}{\mathcal{A}^{\beta}}\right) \bigg \} \left( \xi \mathcal{A}^{-1} -\delta\frac{\mathcal{A}^{\delta-1}}{\sigma^2} \mathcal{I}\left(\frac{v_0}{\mathcal{A}^{\beta}}\right) + v_0 \beta \frac{\mathcal{A}^{\delta - \beta +1}}{\sigma^2}\mathcal{I}'\left(\frac{v_0}{\mathcal{A}^{\beta}}\right)  \right)
    \end{aligned}$
    \item $\begin{aligned}
    \gamma v_0 \frac{\partial f_{v_0}(\mathcal{A})}{\partial v_0} &= \gamma v_0 \frac{\partial}{\partial v_0} \left( v_0 \mathcal{A}^{\xi} \exp \bigg \{ - \frac{\mathcal{A}^{\delta}}{\sigma^2}  \mathcal{I}\left(\frac{v_0}{\mathcal{A}^{\beta}}\right) \bigg \} \right)\\
    &= \gamma v_0 \mathcal{A}^{\xi}  \exp \bigg \{ - \frac{\mathcal{A}^{\delta}}{\sigma^2}  \mathcal{I}\left(\frac{v_0}{\mathcal{A}^{\beta}}\right) \bigg \} \left( 1 - v_0 \frac{\mathcal{A}^{\delta - \beta}}{\sigma^2} \mathcal{I}' \left( \frac{v_0}{\mathcal{A}^{\beta}} \right)\right)
    \end{aligned}$
    \item
    $\begin{aligned}
    \frac{\sigma^2}{2}\frac{\partial^2 f_{v_0}(\mathcal{A})}{\partial v_0^2}=& \frac{\sigma^2}{2} \frac{\partial^2}{\partial v_0^2} \left( v_0 \mathcal{A}^{\xi} \exp \bigg \{ - \frac{\mathcal{A}^{\delta}}{\sigma^2}  \mathcal{I}\left(\frac{v_0}{\mathcal{A}^{\beta}}\right) \bigg \} \right) \\
    =& \frac{\sigma^2}{2}\mathcal{A}^{\xi} \exp \bigg \{-\frac{\mathcal{A}^{\delta}}{\sigma^2}  \mathcal{I}\left(\frac{v_0}{\mathcal{A}^{\beta}}\right) \bigg \} \bigg( -2 \frac{\mathcal{A}^{\delta-\beta}}{\sigma^2} \mathcal{I}' \left( \frac{v_0}{\mathcal{A}^{\beta}} \right) - v_0 \frac{\mathcal{A}^{\delta-2\beta}}{\sigma^2} \mathcal{I}''\left( \frac{v_0}{\mathcal{A}^{\beta}} \right) \\
    &   + v_0 \frac{\mathcal{A}^{2\delta - 2 \beta }}{\sigma^4} \left( \mathcal{I}' \left( \frac{v_0}{\mathcal{A}^{\beta}}\right)\right)^2 \bigg)
    \end{aligned}$
\end{itemize}

\noindent Now putting everything together and deleting the common factors the backward equation \eqref{pdffirstpassageareas} becomes:

\begin{align}
0=&-v_0^{n+1} \xi \mathcal{A}^{-1} + v_0^{n+1} \delta \frac{\mathcal{A}^{\delta-1}}{\sigma^2} \mathcal{I} \left( \frac{v_0}{\mathcal{A}^{\beta}} \right) - v_0^{n+2} \beta \frac{\mathcal{A}^{\delta-\beta+1}}{\sigma^2} \mathcal{I}'\left( \frac{v_0}{\mathcal{A}^{\beta}} \right) -\gamma v_0 + \gamma v_0^2 \frac{\mathcal{A}^{\delta-\beta}}{\sigma^2} \mathcal{I}'\left( \frac{v_0}{\mathcal{A}^{\beta}} \right) \\ \nonumber
& - \mathcal{A}^{\delta-\beta}\mathcal{I}'\left(\frac{v_0}{\mathcal{A}^{\beta}}\right) -\frac{v_0}{2}\mathcal{A}^{\delta-2\beta} \mathcal{I}''\left( \frac{v_0}{\mathcal{A}^{\beta}} \right) + \frac{v_0}{2}\frac{\mathcal{A}^{2\delta - 2\beta}}{\sigma^2}\left( \mathcal{I}'\left( \frac{v_0}{\mathcal{A}^{\beta}} \right) \right)^2
\end{align}

\noindent Now we can perform a change of variables $\omega=\frac{v_0}{\mathcal{A}^{\beta}}$ and massage our equation a little more dividing everything for $\mathcal{A}^{\delta}$:

\begin{align}
0=& -\xi \mathcal{A}^{n\beta + \beta -\delta - 1} + \delta \omega^{n+1} \frac{\mathcal{A}^{n\beta + \beta - 1}}{\sigma^2} \mathcal{I}(\omega) -\beta \omega^{n+2} \frac{\mathcal{A}^{n\beta + \beta - 1}}{\sigma^2} \mathcal{I}'(\omega) - \gamma \omega \mathcal{A}^{\beta - \delta} + \gamma \omega^2 \frac{\mathcal{A}^{\beta}}{\sigma^2} \mathcal{I}'(\omega) \\ \nonumber
& - \mathcal{A}^{-\beta}\mathcal{I}'(\omega) - \frac{\omega}{2} \mathcal{A}^{-\beta} \mathcal{I}''(\omega) + \frac{\omega}{2} \frac{\mathcal{A}^{\delta - \beta}}{\sigma^2} \left( \mathcal{I}'(\omega) \right)^2
\end{align}

\noindent At this point we recall from our initial ansatz that $\beta$ and $\delta$ are critical exponents of an anomalous rate function, hence $\beta, \delta \less 1$, and in particular we can assume without loss of generality in our scaling argument that $\delta \great \beta$. Now, to obtain the rate function $\mathcal{I}(\omega)$ we take the limit $\mathcal{A} \rightarrow \infty$. Note that with this choice taking this limit is equivalent to considering the small noise limit $\sigma \rightarrow 0$, since the leading terms that survive always depend on $\sigma$. Neglecting the subleading contributions in large $\mathcal{A}$ limit, or equivalently in small $\sigma$ limit, we arrive at the equation for the rate function:

\begin{equation}
\frac{1}{\sigma^2} \left( \delta w^{n}\mathcal{A}^{n\beta + \beta -1} \mathcal{I}(w) - \beta w^{n+1} \mathcal{A}^{n\beta + \beta -1} \mathcal{I}'(w) + \gamma w \mathcal{A}^{\beta} \mathcal{I}'(w) + \frac{\mathcal{A}^{\delta -\beta}}{2}  \left(\mathcal{I}'(w) \right)^2 \right) + o\left( 1 \right) = 0
\end{equation}

\noindent Then, imposing the same scaling of $\mathcal{A}$ for all terms of order $\sigma^{-2}$ we obtain the conditions for the scaling exponents $\beta$ and $\delta$, i.e. $n\beta + \beta - 1 = \beta=\delta - \beta$, from which we obtain the values reported in \eqref{scalingexpo}. Then the equation becomes exactly \eqref{ratefunctionequation}. Let us now discuss when its solutions can display an absolute minimum at $\omega=\omega^*>0$. From \eqref{ratefunctionequation} we have $\mathcal{I}'(\omega^*)=0$
only if $\omega^*=0$ or for $\mathcal{I}(\omega^*)=0$, since $\omega^*>0$ we keep $\mathcal{I}(\omega^*)=0$; we notice that $\mathcal{I}(\omega)=0$ for all $\omega$ is a trivial solution of equation \eqref{ratefunctionequation}. Let us consider its derivative with respect to $\omega$:

\begin{equation}
\label{deriv_ratefunctionequation}
\mathcal{I}'(\omega)\mathcal{I}''(\omega) + \mathcal{I}''(\omega)\left( \gamma \omega - \frac{\omega^{n+1}}{n} \right) 
+ \mathcal{I}'(\omega)\left( \gamma - \frac{(n+1)\omega^{n}}{n} \right) 
+\frac{2}{n}\omega^n \mathcal{I}'(\omega) + 2 \omega^{n-1} \mathcal{I}(\omega)= 0.
\end{equation}
Setting $\omega=\omega^*$ since $\mathcal{I}(\omega^*)=0$ and $\mathcal{I}'(\omega^*)=0$ we get $\mathcal{I}''(\omega^*)=0$ or $( \gamma \omega^* - {(\omega^*)^{n+1}}/{n})$. The first solution corresponds to the trivial case $\mathcal{I}(\omega)=0$, the second gives us as expected $\omega^*=\omega_c$. Let us take the derivative of Equation \eqref{deriv_ratefunctionequation}, so we get:

\begin{equation}
\label{deriv_ratefunctionequation}
\mathcal{I}'(\omega)\mathcal{I}'''(\omega) +
\left( \mathcal{I}''(\omega)\right)^2+ 
\mathcal{I}'''(\omega)\left( \gamma \omega - \frac{\omega^{n+1}}{n} \right) 
+ 2 \mathcal{I}''(\omega)\left( \gamma - \omega^{n} \right) 
+ (3-n) \omega^{n-1} \mathcal{I}'(\omega)
+2(n-1)\omega^{n-2} \mathcal{I}(\omega)= 0.
\end{equation}
At $\omega=\omega^*=\omega_c$ we get $\mathcal{I}''(\omega^*)=0$ (the trivial solution) or $\mathcal{I}''(\omega^*)=2 (n-1) \gamma$ which describes the small fluctuations around the noiseless solution $\omega_c$: i.e. $\mathcal{I}(\omega)\sim \gamma(n-1)(\omega-\omega_c)^2$. So Equation \eqref{ratefunctionequation} for the rate function $\mathcal{I}(\omega)$ can be traced back to an already studied Fokker-Planck equation \cite{BarkaiStretchedExp, BarkaiStretchedExp2} for a Brownian particle immersed in a heat bath with a weak binding potential of the same form. To show explicitly the mathematical equivalence between these two problems, it suffices to redefine the variable properly. We start by dividing everything by $\omega^n$:

\begin{equation}
\frac{2}{n}\mathcal{I}(\omega) - \mathcal{I}'(\omega) \left( \frac{2}{n} \omega - \frac{\gamma}{\omega^{n-1}} \right) +\frac{1}{2 \omega^n} \left( \mathcal{I}'(\omega)\right)^2=0
\end{equation}

\noindent Now we only need to define a new variable $z$ such that the quadratic term in the first derivative of the rate function has a pre-factor equal to one, i.e. we seek a change of variable transformation $\omega \rightarrow z$ such that $\frac{1}{2 \omega^n} (\partial_{\omega} \mathcal{I}(\omega))^2 \rightarrow (\partial_z \mathcal{I}(z))^2$, and so using the chain rule we get the Jacobian:

\begin{equation}
\frac{1}{2\omega^n} \left(\frac{\partial \mathcal{I}}{\partial \omega} \right)^2 =  \frac{1}{2\omega^n} \left( \frac{\partial \mathcal{I}}{\partial z} \frac{\partial z}{\partial \omega} \right)^2 = \frac{1}{2\omega^n} \left( \frac{\partial z}{\partial \omega} \right)^2 \left( \frac{\partial \mathcal{I}}{\partial z} \right)^2 \equiv \left(\frac{\partial \mathcal{I}}{\partial z}\right)^2 \ \rightarrow  \ \frac{1}{2\omega^n} \left( \frac{\partial z}{\partial \omega} \right)^2 = 1
\end{equation}

\noindent Solving this simple differential equation, we obtain that the new variable $z$ is:

\begin{equation}
z= \frac{2\sqrt{2}}{n+2} \omega^{\frac{n+2}{2}} \ \rightarrow \  \omega=\left(\frac{n+2}{2\sqrt{2}} z\right)^{\frac{2}{n+2}}
\end{equation}

\noindent Now we know the expression of $z$ and we can calculate the derivative of the new rate function $\mathcal{I}(z)$:

\begin{equation}
\frac{\partial \mathcal{I}}{\partial w} = \frac{\partial z}{\partial \omega} \frac{\partial \mathcal{I}}{\partial z} = \sqrt{2} \omega^{\frac{n}{2}} \frac{\partial \mathcal{I}}{\partial z} = \sqrt{2}\left(\frac{n+2}{2\sqrt{2}}z\right)^{\frac{n}{n+2}} \frac{\partial \mathcal{I}}{\partial z}  
\end{equation}

\noindent Putting everything together in \eqref{ratefunctionequation}, we find that the new backward equation for $\mathcal{I}(z)$ becomes:

\begin{equation}
\frac{2}{n}\mathcal{I}(z) - \mathcal{I}'(z)\left( \frac{n+2}{n} z - \gamma \sqrt{2}\left( \frac{n+2}{2\sqrt{2}} z \right)^{-\frac{n-2}{n+2}} \right) + (\mathcal{I}'(z))^2 = 0 
\end{equation}

\noindent We can identify the power term of the effective weak potential by noting that $\frac{n-2}{n+2} = 1 - \frac{4}{n+2}$, defining therefore $\alpha=\frac{4}{n-2}$, which is the exponent of the weak power potential $V(x) \sim V_0 x^{\alpha}$. At this point, the equation written in terms of $\alpha$ is:

\begin{equation}
\label{backwardeqnn}
\frac{\alpha \mathcal{I}(z)}{2-\alpha} - \mathcal{I}'(z)\left( - \gamma\left( \frac{\sqrt{2}}{\alpha} \right)^{\alpha} \frac{\alpha}{z^{1-\alpha}} +\frac{z}{2-\alpha} \right) + (\mathcal{I}'(z))^2 = 0
\end{equation}

\noindent An interesting observation is that this final equation for the rate function \eqref{backwardeqnn} is actually the same equation for the rate function equation in \cite{BarkaiStretchedExp, BarkaiStretchedExp2}, and comparing this expression with that studied in that work, one can easily realise that the dimensionless energy ratio connecting the potential to the heat bath is given by $\frac{V_0}{k_B T} = \gamma\left( \frac{\sqrt{2}}{\alpha} \right)^{\alpha}$, and there is also a minus sign difference in the term in $z^{\alpha-1}$. This difference is due to the fact that the solution of this rate function equation is shifted by the potential term with respect to the one in \cite{BarkaiStretchedExp, BarkaiStretchedExp2} given by $\mathcal{I}(z) = \mathcal{I}_{forward}(z)- \frac{V_0}{k_B T}z^{\alpha}$. This shift term in the potential is related to the fact that in the cited work the authors start by using a forward formulation of the Fokker-Planck equation for the PDF of the weak-binding system, instead in our work we start with a backward Fokker-Planck equation for the PDF $f_{v_0}(\mathcal{A})$, since we are looking for a first passage PDF in which we need to fix the final condition and integrate with respect to the initial condition $v_0$. We know that the forward formulation in the limit of $z \rightarrow \infty$ the rate function equation \label{backward} (corrected with the plus sign in the term in $z^{-1+\alpha}$) has two possible solutions, namely a Boltzmann Gibbs solution at equilibrium $\mathcal{I}_{BG forward}(z)$, which coincides with the potential term $\frac{V_0}{k_B T}z^{\alpha}$, and a diffusive solution $\mathcal{I}_{D}(z)$. In our backward formulation, instead, the Boltzmann-Gibbs solution coincides with the trivial one, i.e. $\mathcal{I}_{BG}(z)=\mathcal{I}_{BG forward}(z) - \frac{V_0}{k_B T} = 0$ due to the shift of the potential. Conversely, in the case of the diffusive term $\mathcal{I}_{D}(z) = \mathcal{I}_{D forward}(z) - \frac{V_0}{k_B T}$ we can still obtain the non-trivial value $I_D(z) \sim \frac{2}{(n+2)^2} \omega^{n+2}$, and this fact explains the mismatch between the rate function plot in Figure 1 of \cite{BarkaiStretchedExp, BarkaiStretchedExp2} and our plot in Figure \ref{fig:ratefunction_analytical}.
\noindent We also know that the value for $z \rightarrow 0$ of the diffusive solution, that is the same in both the backward and the forward formulation, obtained using a WKB method \cite{Bender}, is given by:

\begin{equation}
I_{D}(0) = \frac{\left( \sqrt{\pi} \left( \alpha \gamma\left( \frac{\sqrt{2}}{\alpha} \right)^{\alpha}  \right)^{\frac{1}{1-\alpha}}\frac{\Gamma\left(\frac{\alpha}{2-2\alpha}\right)}{\Gamma\left( \frac{1}{2-2\alpha} \right)} \right)^{1-\nu}}{\nu^{\nu}(1-\nu)^{1-\nu}} \ \text{with $\nu=\frac{\alpha}{2-\alpha}$}
\end{equation}

\noindent At this point, knowing the value of $\alpha$ in terms of $n$, one can obtain exactly the exponential pre-factor $c_n$ \eqref{coeffcn} already calculated in \cite{TouchetteInst, MeersonInst, NaftaliInst} with a different approach based on a path integral formulation. Regarding the calculation of the exponent $\xi$ for the subleading power term in the $\mathcal{A}\rightarrow 0$ solution of \eqref{pdfAgeneral}, it is possible to obtain it in a simple way in the limit of $\omega \rightarrow \infty$ by considering the diffusive solution. We know that the first passage PDF $f_{v_0}(\mathcal{A})$ of Brownian motion must have a power prefactor term \cite{MajumdarFPA}, so a simple argument for fixing its critical exponent is simply imposing the normalization of the PDF. Then, we get:  

\begin{equation}
\int_0^{\infty} v_0 \mathcal{A}^{\xi} \exp \bigg \{-\frac{2}{(n+2)^2 \sigma^2} \frac{v_0^{n+2}}{\mathcal{A}}\bigg \} d\mathcal{A} = 1 \ \rightarrow \ \xi = -1 -\frac{1}{n+2}
\end{equation}

\bibliography{BJ_biblio}

\end{document}